%% file: paper.tex
\def\maintitle{Systemization of Pluggable Transports for Censorship Resistance}
\def\papertitle{\maintitle}

\documentclass[USenglish,oneside,twocolumn]{article}
\PassOptionsToPackage{hidelinks}{hyperref}
\usepackage[big]{dgruyter_NEW}
\usepackage{times}
\usepackage{url}
\usepackage{array,multirow}
\usepackage{bigstrut}
\usepackage{xspace}
\usepackage[usenames,dvipsnames]{color}
\usepackage[font={footnotesize}]{caption}
\usepackage{subcaption}
\usepackage{amsmath}
\usepackage{mathtools}

\usepackage{prettyref}
\newrefformat{sec}{Section~\ref{#1}}
\newrefformat{tab}{Table~\ref{#1}}
\newrefformat{fig}{Figure~\ref{#1}}
\newrefformat{cha}{Chapter~\ref{#1}}
\newrefformat{app}{Appendix~\ref{#1}}
\newrefformat{eqa}{Equation~\ref{#1}}

\usepackage{comment}
\usepackage{ wasysym }
\usepackage{ amssymb }
\usepackage{ lscape }
\usepackage{ upgreek }
\usepackage{color}

\makeatletter
\newcommand{\ie}{i.e.\@\xspace}
\newcommand{\eg}{e.g.\@\xspace}

\newcommand{\etal}{\textit{et al.\@\xspace}}
\newcommand{\detal}{\textit{et al.}}

\newcommand\etalcite[1]{\etal~\cite{#1}}
\makeatother

\usepackage{booktabs}
\newcolumntype{P}[1]{>{\raggedright\arraybackslash}p{#1}}

\usepackage{hyperref}

\definecolor{linkcol}{rgb}{0,0,1}
\definecolor{citecol}{rgb}{0,0.5,0}
\definecolor{urlcol}{rgb}{0.3,0,0}

\renewcommand{\and}{\hspace{5mm}}

\def\first{({\it i})\xspace}
\def\second{({\it ii})\xspace}
\def\third{({\it iii})\xspace}
\def\fourth{({\it iv})\xspace}

\newcommand\mypara[1]{\noindent\textbf{#1.}\newline}

\newcolumntype{R}[2]{%
    >{\adjustbox{angle=#1,lap=\width-(#2)}\bgroup}%
    l%
    <{\egroup}%
}
\def\censorspoofer{{CensorSpoofer}\xspace}
\def\scramblesuit{{ScrambleSuit}\xspace}
\def\stegotorus{{StegoTorus}\xspace}
\def\meek{{Meek}\xspace}
\def\dust{{Dust}\xspace}
\def\obfsTwo{{obfs2}\xspace}
\def\obfsThree{{obfs3}\xspace}
\def\obfsFour{{obfs4}\xspace}
\def\flashproxy{{Flashproxy}\xspace}
\def\fte{{FTE}\xspace}

\def\skypemorph{{SkypeMorph}\xspace}
\def\freewave{{Freewave}\xspace}
\def\facet{{Facet}\xspace}
\def\sweet{{SWEET}\xspace}
\def\mailmyweb{{MailMyWeb}\xspace}
\def\telex{{Telex}\xspace}
\def\cirripede{{Cirrepede}\xspace}
\def\curveball{{Curveball}\xspace}
\def\tapdance{{TapDance}\xspace}

\def\cloudtransport{{CloudTransport}\xspace}
\def\transteg{{TransTeg}\xspace}
\def\trist{{TRIST}\xspace}
\def\infranet{{Infranet}\xspace}
\def\collage{{Collage}\xspace}
\def\miab{{MIAB}\xspace}
\def\oss{{OSS}\xspace}

\def\defiance{{Defiance}\xspace}

\def\mesg-stream-encrypt{{MSE}\xspace}
\def\id-based-crypto-tagging{{IBS}\xspace}

\def\foe{{FOE}\xspace}
\def\silentknock{{SilentKnock}\xspace}
\def\bridgespa{{BridgeSPA}\xspace}
\def\spator{{SPATor}\xspace}
\def\vpngate{{VPN-Gate}\xspace}
\def\jumpbox{{JumpBox}\xspace}
\def\marionette{{Marionette}\xspace}
\def\rook{{Rook}\xspace}
\def\castle{{Castle}\xspace}
\def\gohop{{GoHop}\xspace}
\def\keyspacehopping{{Keyspace-Hopping}\xspace}

\def\pt{{LC}\xspace}
\def\pts{{LCs}\xspace}

\def\others{{Miscellaneous}\xspace}

\def\cenSer{{CEN.SER}\xspace}
\def\cenCli{{CEN.CLI}\xspace}
\def\corCon{{COR.CON}\xspace}
\def\corRou{{COR.ROU}\xspace}
\def\blkRou{{BLK.ROU}\xspace}

\def\corSem{{COR.SEM}\xspace}

\def\fprLen{{FPR.LEN}\xspace}
\def\fprTim{{FPR.TIM}\xspace}
\def\fprSem{{FPR.SEM}\xspace}
\def\fprRou{{FPR.ROU}\xspace}
\def\fprCon{{FPR.CON}\xspace}
\def\degPer{{DEG.PER}\xspace}

\def\npapers{41\xspace}

\def\SI{\textsl{SI}\xspace}
\def\ENC{\textsl{ENC}\xspace}
\def\MUX{\textsl{MUX}\xspace}
\def\OBF{\textsl{OBF}\xspace}
\def\TIM-LEN{\textsl{TIM-LEN}\xspace}
\def\TRN{\textsl{TRN}\xspace}

\def\session_init{\textsl{Session Initialisation}\xspace}
\def\encrypt{\textsl{Encryption}\xspace}
\def\contentObfs{\textsl{Content Obfuscation}\xspace}
\def\conObfs{\textsl{Content Obfuscation}\xspace}
\def\timeObfs{\textsl{Timing Obfuscation}\xspace}
\def\lenObfs{\textsl{Length Obfuscation}\xspace}
\def\transport{\textsl{Transport}\xspace}
\def\multiplex{\textsl{Multiplexing}\xspace}

\newcolumntype{?}{!{\vrule width 1pt}}
\definecolor{myred}{RGB}{215,25,28}
\definecolor{myyellow}{RGB}{253,174,97}
\definecolor{myorange}{RGB}{253,174,97}
\definecolor{mygreen}{RGB}{215,25,28}
\definecolor{myblue}{RGB}{43,131,186}
\newcommand{\si}{\textcolor{myred}{$\blacktriangle$}} %
\newcommand{\enc}{\textcolor{myred}{$\spadesuit$}}	%
\newcommand{\mux}{\textcolor{myyellow}{$\clubsuit$}}	%
\newcommand{\cobf}{\textcolor{myorange}{$\blacksquare$}}	%
\newcommand{\tobf}{\textcolor{mygreen}{$\blacklozenge$}}	%
\newcommand{\lobf}{\textcolor{mygreen}{$\blacktriangledown$}}	%
\newcommand{\trn}{\textcolor{myblue}{$\bigstar$}}	%
\newcommand\mycomment[1]{#1}
\newcommand\sbk[1]{\mycomment{\textit{}}}

\newcommand\ls[1]{\mycomment{\textit{\textbf{}}}}

\begin{document}

\title{\papertitle}

\author*[1]{Sheharbano Khattak}
\affil[1]{University of Cambridge, \mbox{Sheharbano.Khattak@cl.cam.ac.uk}}

\author[2]{Laurent Simon}
\affil[2]{University of Cambridge, \mbox{Laurent.Simon@cl.cam.ac.uk}}

\author[3]{Steven J. Murdoch}
\affil[3]{University College London, \mbox{s.murdoch@ucl.ac.uk}}

\input{abstract}

\maketitle

\input{intro}

\input{background}
\input{attack-tree}
\input{link-obfs}
\input{ip-resist}

\input{flow-resist}

\input{comp-resist}

\input{discussion}
\input{conclusion}

\begin{acknowledgement}
We appreciate the valuable feedback we received from the anonymous reviewers. Steven J. Murdoch and Sheharbano Khattak were supported by The Royal Society [grant number UF110392]; Engineering and Physical Sciences Research Council [grant number EP/L003406/1]. Laurent Simon was supported by Samsung SERI and Thales e-security.
\end{acknowledgement}

\bibliographystyle{ieeetr}
\bibliography{paper}

\end{document}

%% file: abstract.tex
\begin{abstract} 
{An increasing number of countries implement Internet censorship at
different scales and for a variety of reasons. In particular, the link
between the censored client and entry point to the uncensored network
is a frequent target of censorship due to the ease with which a
nation-state censor can control it. A number of censorship
resistance systems have been developed thus far to help circumvent
blocking on this link, which we refer to as \textsl{link circumvention
systems} (\pts). The variety and profusion of attack vectors available
to a censor has led to an arms race, leading to a dramatic speed of
evolution of \pts.  Despite their inherent complexity and the breadth
of work in this area, there is no systematic way to evaluate link
circumvention systems and compare them against each other. In this
paper, we \first sketch an attack model to comprehensively explore a
censor's capabilities, \second present an abstract model of a 
\pt, a system that helps a censored client
communicate with a server over the Internet while resisting
censorship, \third describe an evaluation stack that underscores a
layered approach to evaluate \pts, and \fourth systemize and evaluate
existing censorship resistance systems that provide link
circumvention. We highlight open challenges in the evaluation and
development of \pts and discuss possible mitigations.\\
\\
\textit{Content from this paper was published in Proceedings on Privacy Enhancing Technologies (PoPETS), Volume 2016, Issue 4 (July 2016) as "SoK: Making Sense of Censorship Resistance Systems" by Sheharbano Khattak, Tariq Elahi, Laurent Simon, Colleen M. Swanson, Steven J. Murdoch and Ian Goldberg. DOI: \href{http://dx.doi.org/10.1515/popets-2016-0028}{10.1515/popets-2016-0028}}} 
\end{abstract}

%% file: intro.tex
\section{Introduction} 

\label{intro} 

\begin{figure} \centering
\includegraphics[width=3.0in]{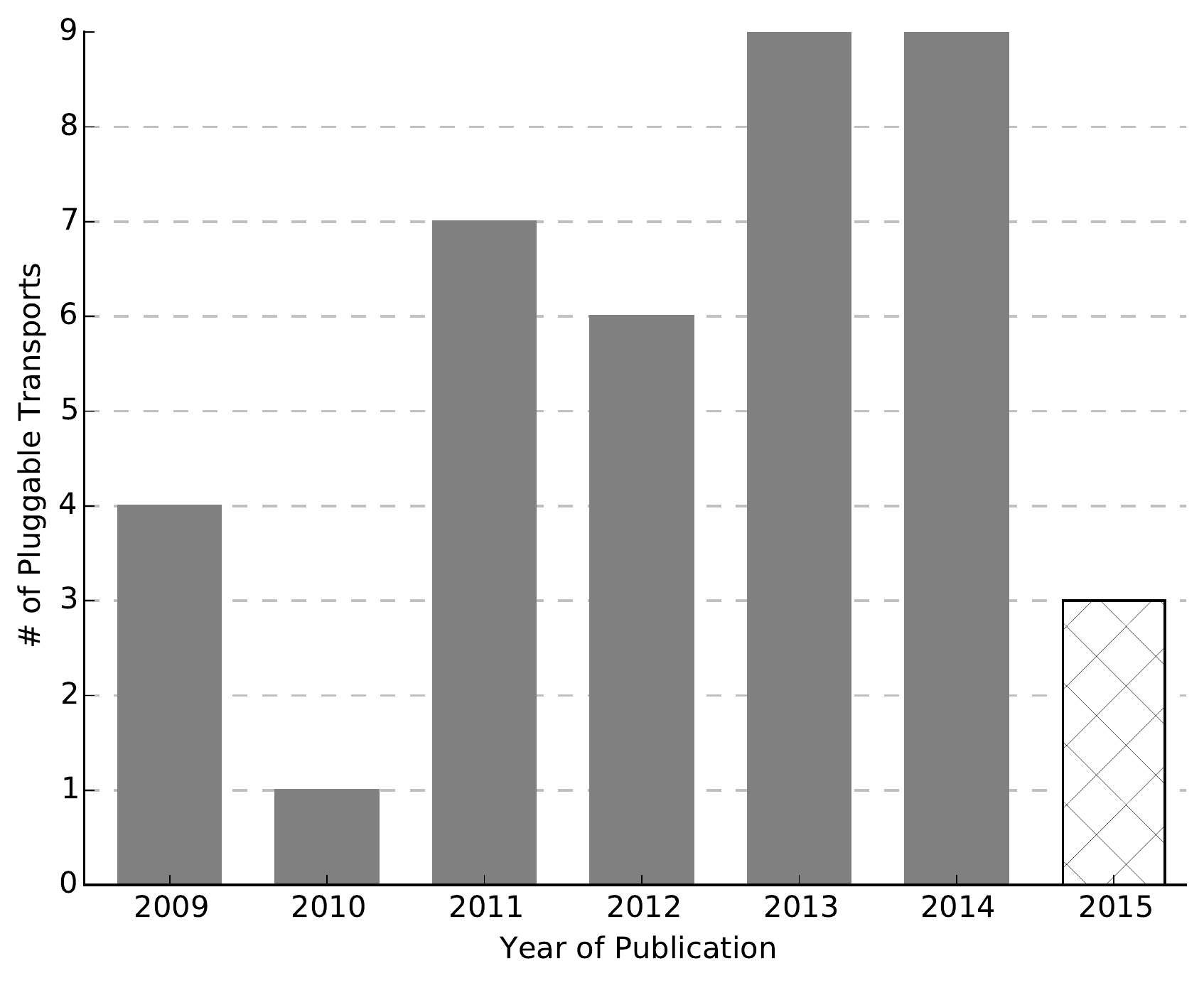}

{\caption{ Surveyed systems (literature and implementation) from the
last five years related to censorship resistance systems with a focus on link circumvention.}
\label{fig:PT-timeline}}

\end{figure}

As the Internet becomes an increasingly important means to engage in
civil society, those who wish to control the flow of information are
turning to measures to suppress speech which they consider
undesirable. While blocking can take place at a number of points on
a network, the link between the censored client and entry point to
uncensored part of the network has been a frequent target as the
censor is typically a powerful nation-state adversary in control of
network infrastructure within the censored region. A number of
censorship resistance systems have emerged to help bypass such blocks
which we call link circumvention systems or \pts (the scope of this
systemization is limited to systems that resist censorship on the
link). Due to the diversity of censorship mechanisms across different
jurisdictions and their evolution over time, there is no one approach
which is optimally efficient and resistant to all censors.
Consequently, an arms race has developed resulting in the evolution of
\pts to have dramatically sped up with resistance schemes becoming
increasingly complex. This has been captured in
\prettyref{fig:PT-timeline} which shows work in this area over the
last five years as per our survey.  

Despite the multitude of \pts that have
been developed so far, there is no systematic way to evaluate them.
The area lacks a comprehensive attack model of the censor
which has led to systems assuming arbitrary threat models, some of
which are misaligned with how real censors
operate~\cite{sadia2015hotpets}. As a result of the disparate attack
vectors these systems protect against, it is hard to assess the scope
of circumvention offered by a system in isolation as well as in
comparison with others.

We systematize and evaluate existing work by conducting a comprehensive
survey of censorship resistance systems that offer link circumvention
(we identify \npapers such systems). To understand effectiveness of
various \pts under different censorship
scenarios, we sketch a comprehensive attack model to understand a
censor's capabilities. Next we develop an abstract model of a link
circumvention system that enables a client application in censored
region to communicate with a server application over the Internet,
even though direct connections are blocked.\footnote{An instance of
our abstract link circumvention (\pt) model is the de facto API for
anonymous communication systems to integrate with censorship
resistance schemes~\cite{pt_spec_tor}, however, we purposely maintain
a broad focus to accommodate \pts that have not been written strictly
as a Pluggable Transport but can fit the broad model with minor
adaptation.} We then present an evaluation stack that highlights
capabilities of a \pt in terms of the mechanisms employed to resist
different censorship vectors. 

\textsl{Related Work:} Existing work  systematizes censorship and circumvention systems in a broader context.
Elahi and Goldberg~\cite{elahi2012cordon} present a taxonomy of
censorship resistance strategies for different types of censors (\eg
ISP, government) with the decision-making process based on their
resources, capabilities, limitations and utility.  Tschantz
\detal~\cite{Tschantz2014} argue that the evaluation of circumvention
tools should be based on economic models of censorship.
K{\"o}psell~\etalcite{kopsell2004achieve} present a classification of
blocking techniques based on the communication layer involved, the
content of communication (\eg images, web) and metadata of the
communication (\eg IP addresses of participants, time of the
communication, protocols involved).
Perng~\etalcite{perng2005censorship} classify circumvention systems based on the
technical primitives and principles they build upon.
Leberknight~\etalcite{leberknight2012taxonomy} survey the social,
political and technical aspects that underpin censorship.  They also
propose metrics that quantify their efficacy, \ie scale, cost and
granularity. Our work has a special focus on the link circumvention aspect
of censorship resistance, and extends previous work by using a
comprehensive attack model and evaluation stack to systematize link
circumvention systems.

In this systemization of knowledge paper, we make the following
contributions: 

\begin{itemize}

\item present an attack model from a censor's perspective that
captures capabilities of a censor and the diversity of censorship
mechanisms (Section~\ref{attack-tree}).

\item develop an abstract model of a link circumvention system that
facilitates communication between a censored client and server in a
censorship resistant fashion (Section~\ref{sec:pt-model}). 

\item present an evaluation stack to assess scope of circumvention
offered by various link circumvention systems
(Section~\ref{sec:pt-model}).

\item survey \npapers link circumvention systems and evaluate them
using the proposed attack model and evaluation stack. We identify
broad categories of resistance systems based on the attack path(s)
that these seek to protect (Sections~\ref{ip-resist},
\ref{flow-resist} and \ref{comp-resist}).  

\item discuss open challenges in the evaluation of link circumvention
schemes and discuss possible solutions to mitigate these
(Section~\ref{sec:discussion}). 

\end{itemize}

%% file: background.tex
\section{Background} \label{background}

\begin{figure*}[t!]
    \centering
    \begin{subfigure}[t]{1.0\textwidth}
        \centering
        \includegraphics[width=5.0in]{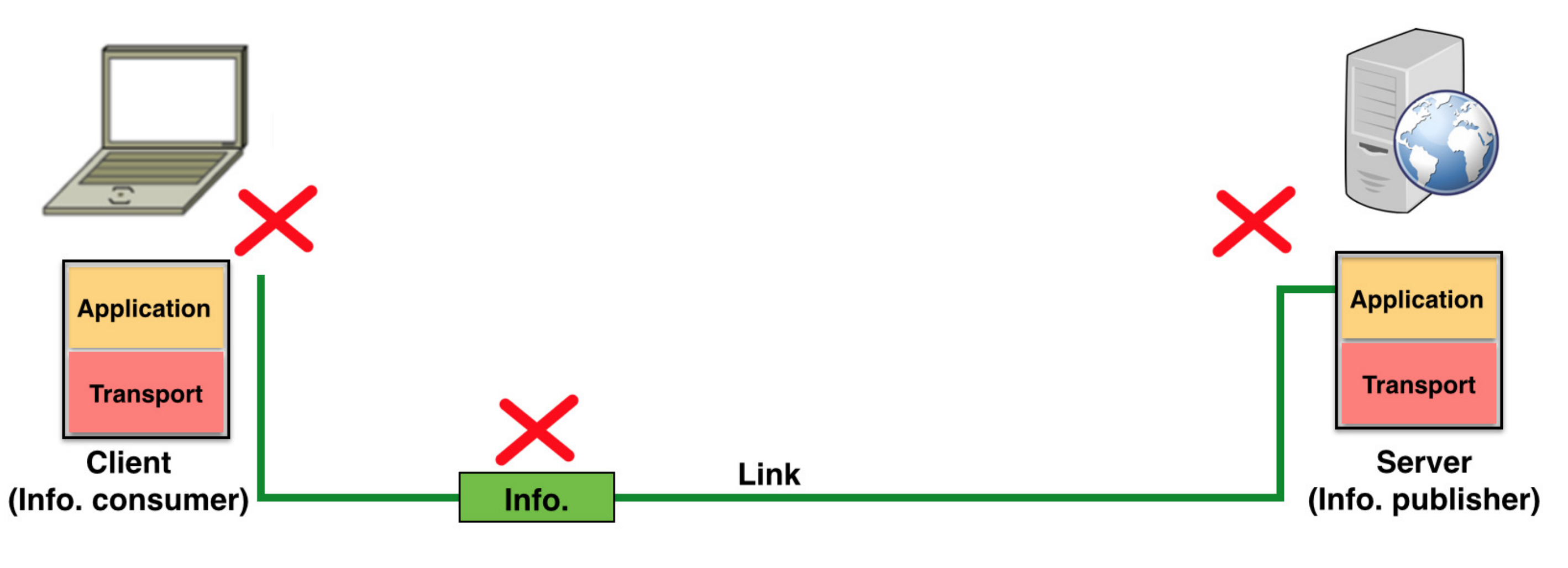}
        \caption{A typical information retrieval process involves an information consumer (the client), an information publisher (the server), and a link that connects the two. A censor's goal is to disrupt dissemination of information either by corrupting it, or by hindering its access or publication (indicated by red crosses in the figure).}
	\label{fig:nw-info}
    \end{subfigure}%

    \begin{subfigure}[t]{1.0\textwidth}
        \centering
        \includegraphics[width=5.0in]{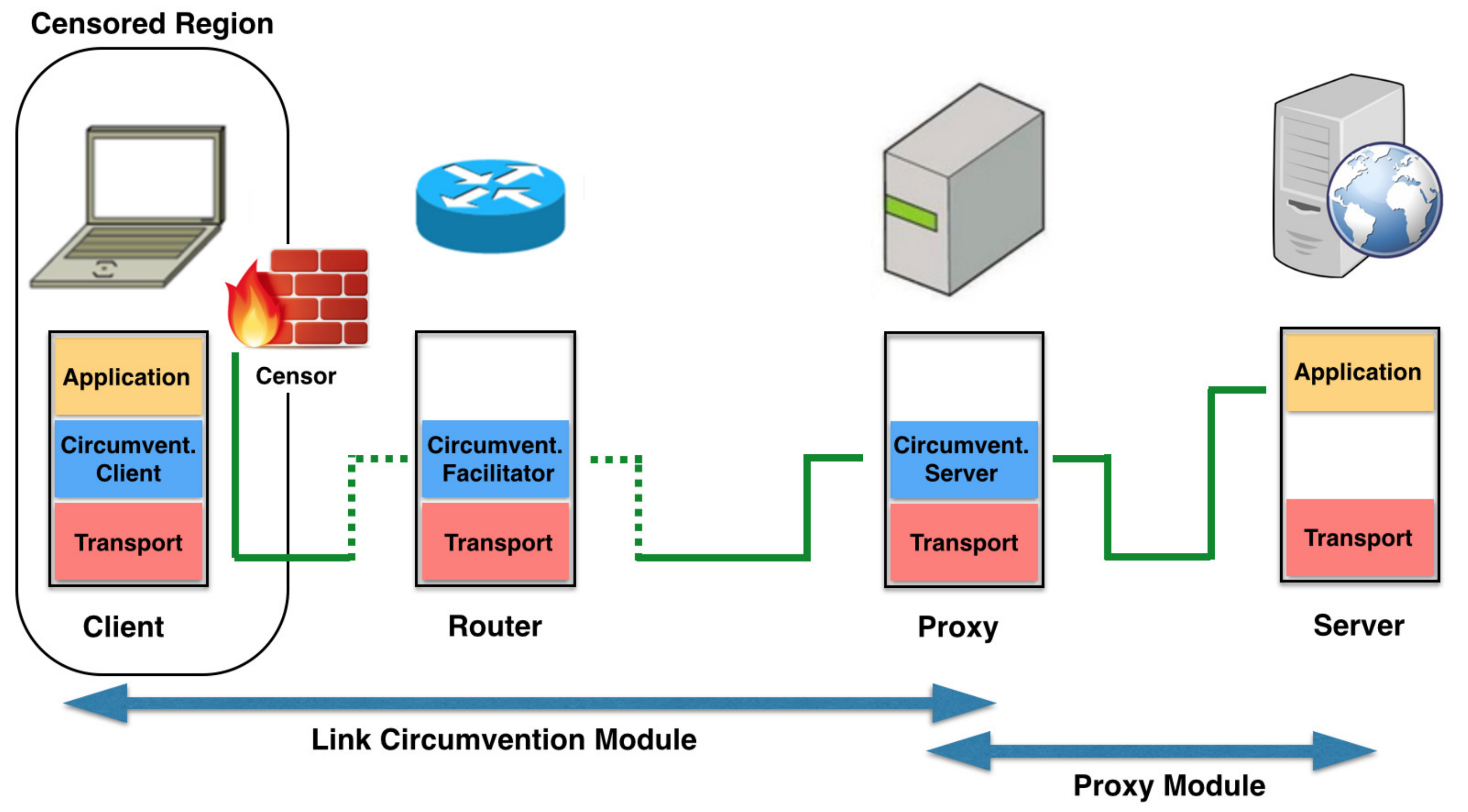}
        \caption{Circumvention software provides unfettered information retrieval despite censorship and may encompass any part of the information network. The process is typically facilitated by a \textsl{proxy} that relays traffic between a \textsl{client} in the censored region and an external \textsl{server}, and effectively divides the channel into two portions which circumvention tools handle separately via different modules: \first client to proxy (\textsl{link circumvention module}), and \second proxy to server (\textsl{proxy module}). The \textsl{circumvention client} allows the client application to construct a communication channel to the server application through a \textsl{circumvention server}, which can optionally be facilitated by an intermediate device (\textsl{circumvention facilitator}).}
      \label{fig:nw-info-circumvent}
    \end{subfigure}
    \captionsetup{justification=centering}
    \caption{Information retrieval (a) without censorship, and (b) with censorship, aided by circumvention software.}
\end{figure*}

In its simplest form, an information dissemination network comprises
an information consumer (the client), an information publisher (the
server) and a link that connects the two (\prettyref{fig:nw-info}). A
censor's goal is to disrupt information dissemination. This can be
done by directly targeting the information (through corruption,
insertion of false information, deletion or modification), or by
impairing access or publication of information.  
 
A censorship resistance system thwarts a censor's attempts to corrupt
information, or its access or publication. While a censorship
resistance system can encompass any part of the information network,
in reality most systems offer circumvention on the link connecting
information endpoints due to its flashpoint status in the censorship
arms race. Link censorship is more prevalent because it is
less intrusive, and convenient (the communication infrastructure is
typically under the censor's direct control). 

For circumvention on the information link, it is common to employ a
proxy, an intermediate unblocked system that relays traffic back and
forth between the client and server. A proxy divides the link between
client and server into two distinct portions: \first client to proxy
(within censored region), and \second proxy to server (outside
censored region). This has been illustrated in
\prettyref{fig:nw-info-circumvent}. Resistance systems increasingly treat these
two portions separately (via \textsl{link circumvention module} and
\textsl{proxy module}, respectively) as these lend themselves to
different design, implementation, and software distribution practices.

The \textsl{proxy module}, being in uncensored region, may simply provide
access to server, without offering any additional security properties
and could be simply implemented as a HTTP or SOCKS proxy, or as a VPN.
Alternatively, the proxy module may be an anonymity system like
Tor~\cite{Tor} which not only provides access to the server but also
prevents attackers from being able to identify which user is accessing
which resource. In contrast, \textsl{link circumvention} being in the line of
fire is a rapidly evolving but less mature area. In this paper, our
focus is the link circumvention role of censorship resistance systems.
Any reference to censorship resistance systems henceforth should be
perceived within the scope of link circumvention.  

%% file: attack-tree.tex
\section{A Censor's Attack Model} 

\label{attack-tree}

\begin{figure*} \centering
\includegraphics[width=17cm]{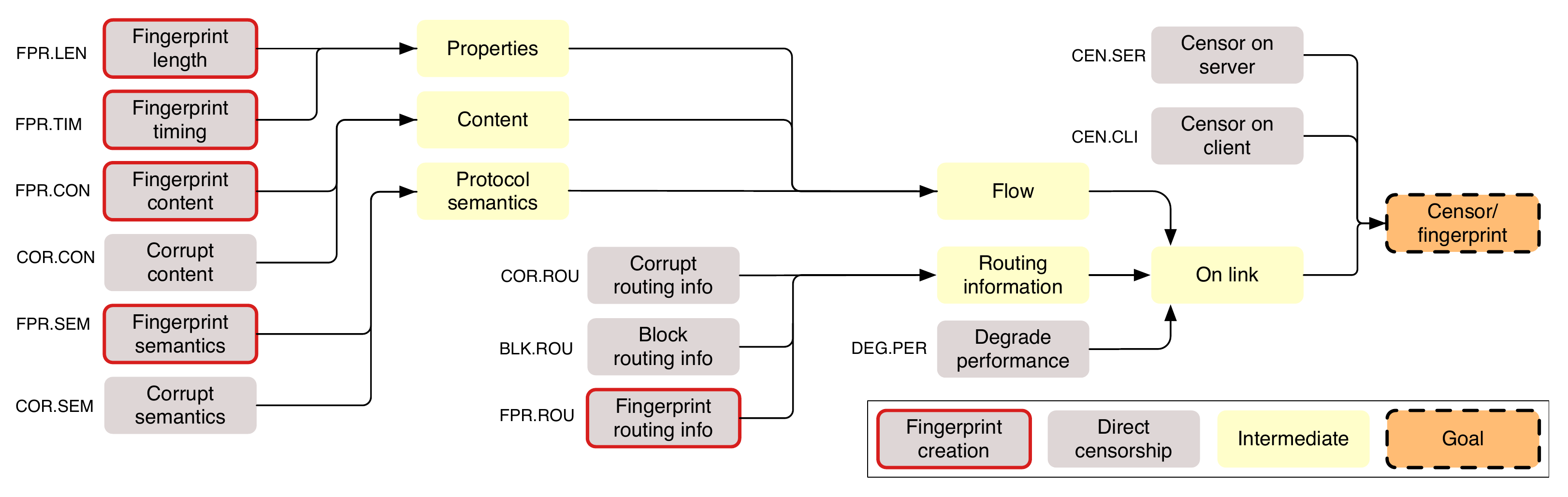}

\caption{\label{fig:attack-tree}Censor's attack model, showing both
direct censorship (information corruption, or disabling access or publication) and
indirect censorship (fingerprinting to develop and improve features for direct censorship)} \end{figure*}

To understand the scope of circumvention offered by various tools, it
is important to first understand the attack surface available to a
censor (illustrated in \prettyref{fig:attack-tree}). While the link
between client and server remains the focus of this paper, for the
sake of completeness, we also include client and server-based
censorship in the attack model. The model captures both direct and
indirect forms of censorship.  While direct censorship immediately
targets information dissemination, indirect censorship (or
\textsl{fingerprinting}) provides input to aid the censor's blocking
decision. For example, a censor can identify a protocol by destination
port (fingerprinting) and follow up by blocking IP addresses in the
associated flows (direct censorship). For convenience, through the
rest of this paper we refer to shorthands associated with direct
censorship and fingerprinting activities described in
Figure~\ref{fig:attack-tree}. We now discuss various censorship
scenarios grouped under direct censorship and fingerprinting.

\subsection{Direct Censorship}

\mypara{Censor on Client (\cenCli)} Client-side censorship can take
place through direct or discrete (facilitated by malware or insider
attacks) installation of surveillance software. This may lead to
corruption of information as well as access disruption. China's Green
Dam, a filtering software product purported to prevent children from
harmful Internet content, was mandated to be installed on all new
Chinese computers in 2009~\cite{greendam}. The software was found to
be far more intrusive than officially portrayed, blocking access to a
large blacklist of websites in diverse categories, and monitored and
disrupted operation of various programs if found to be engaging in
censored activity.  TOM-Skype, a joint venture between a Chinese
telephony company TOM Online and Skype Limited, is a Voice-over-IP
(VoIP) chat client program that uses a list of keywords to censor chat
messages in either direction~\cite{TomSkype}.

\mypara{Censor on Server (\cenSer)} A censor can install software on
the server-side to corrupt the information being published or disrupt
the publication process. A number of studies investigate Chinese
government's censorship of posts on the national microblogging site
Sina Weibo.  Bamman \detal~\cite{Bamman12} analyze three months of
Weibo data and find that 16\% of politically-driven content is
deleted.  Zhu \detal~\cite{Zhu2013} note that Weibo's user-generated
content is mainly removed during the hour following the post with
$\sim$30\% of removals occurring within 30 minutes and $\sim$90\%
within 24 hours.  Another study observes posts from politically active
Weibo users over 44 days and finds that censorship varies across
topics, with the highest deletion rate culminating at 82\%. They
further note the use of \textsl{morphs}--adapted variants of words to
avoid keyword-based censorship.  Weiboscope~\cite{Weiboscope}, a data
collection, image aggregation and visualization tool, makes censored
Sina Weibo posts by a set of Chinese microbloggers publicly available. 

\mypara{Degrade Performance (\degPer)} Network performance degradation
is a soft form of censorship that diminishes access to information
while at the same time affording deniability to the censor.
Anderson~\cite{Anderson2013} uses a set of diagnostics data (such as
network congestion, packet loss, latency) to study the use of
throttling of Internet connectivity in Iran between January 2010 and
2013. He  uncovers two extended periods with a 77\% and 69\% decrease
in download throughput respectively; as well as eight to nine shorter
periods. These often coincide with holidays, protest events,
international political turmoils and important anniversaries, and are
sometimes corroborated by overt filtering of online services or
jamming of international broadcast television.

\mypara{Block Routing Information (\blkRou)} Censorship can leverage
items of a connection tuple (source IP address, source port,
destination IP address and destination port) to disable information
access or publication. The block can continue for a short period of
time to create a chilling effect and encourage self-censorship on part
of the client.  One study notes that the Great Firewall of China (GFW)
blocks communication from a client IP address to a destination IP
address and port combination for 90 seconds after observing
`objectionable' activity over that flow~\cite{gfw-review:httpurl}.
The GFW has been reported to drop packets originating from Tor bridges
based on both source IP address and source port to minimize collateral
damage~\cite{Winter2012}.

\mypara{Corrupt Routing Information (\corRou)} A censor can disrupt
access by corrupting information that supports correct routing of
packets. This can be done by changing routing entries on an
intermediate censor-controlled router, or by manipulating information
that supports the routing process, \eg BGP hijacking and DNS
manipulation. The Border Gateway Protocol (BGP) is the de facto protocol
for inter-AS routing. A censor can block a network's connectivity to
the Internet by withdrawing previously advertised network prefixes or
re-advertising them with different properties (rogue BGP route
advertisements). A number of countries have attempted to effect
complete or partial Internet outages in recent years by withdrawing
their networks in the Internet's global routing table
(Egypt~\cite{egypt-outage}, Libya~\cite{libya-outage},
Sudan~\cite{sudan-outage}, Myanmar~\cite{myanmar-outage}). DNS is
another vital service that maps names given to different Internet
resources to IP addresses. The hierarchical distributed nature of DNS
makes it vulnerable to censorship. Typical forms of DNS manipulation
involve redirecting DNS queries for blacklisted domain names to a
censor-controlled IP address (DNS redirection or poisoning), a
non-existent IP address (DNS blackholing) or by simply dropping DNS
responses for blacklisted domains. China's injection of forged DNS
responses to queries for blocked domain names is well known, and
causes large scale collateral damage by applying the same censorship
policy to outside traffic that traverses Chinese
links~\cite{Anonymous12}.

\mypara{Corrupt Flow Content (\corCon)} A censor can compromise
information or disrupt access by corrupting flow content (\corCon),
\ie the application layer payload. For example, a censor can inject
HTTP 404 Not Found message in response to requests for censored
content and drop the original response, or modify the HTML page in the
body of an HTTP response.

\mypara{Corrupt Protocol Semantics (\corSem)} A censor can corrupt
information or disrupt access by manipulating protocol semantics
(\corSem). A censor can leverage knowledge of protocol specification
to induce disruption on a flow (for example, injecting forged TCP
reset packets into a flow will cause both endpoints to tear down the
connection). 

\subsection{Fingerprinting} \label{fingerprint}

\mypara{Fingerprint Routing Information (\fprRou)} A flow can be associated with a
protocol based on items of the connection tuple. The destination port is a
typical target of censorship (\eg 80 for HTTP). Flows addressed to IP
addresses known to be associated with a blocked service can be
disrupted by implication.  Flow fingerprinting of this kind can form
part of a multi-stage censorship policy, possibly followed by a
blocking step.  Clayton examines the hybrid two-stage censorship
system \textsl{CleanFeed} deployed by British ISP, BT. In the first
stage, it redirects suspicious traffic (based on destination IP and
port) to an HTTP proxy. In the next stage it performs content
filtering on the redirected traffic and returns an error message if
requested content is in the Internet Watch Foundation (IWF)
list~\cite{Clayton2006}.

\mypara{Fingerprint Content (\fprCon)} Flows can be fingerprinted by checking for
the presence of protocol-specific strings, blacklisted keywords,
domain names and HTTP hosts etc.  A number of deep packet inspection (DPI) boxes can perform
regex-based traffic classification~\cite{l7-filter,bro,snort,nDPI},
however it remains unclear what are the true costs of performing
DPI at scale~\cite{Dainotti2012,Sommer2010}.
Alternatively, flows can be fingerprinted based on some property of
the content being carried. For example, a censor that does not allow
encrypted content can block flows where content has high
entropy~\cite{Dorfinger2011}.

\mypara{Fingerprint Flow Properties (\fprLen and \fprTim)} A censor
can fingerprint a protocol by creating its statistical model based on
flow features such as packet length, and timing-related features
(inter-arrival times, burstiness etc.). Once a model has been derived,
a censor can fingerprint flows based on their resemblance or deviation
from this model~\cite{Bernaille2007,Wright2006}.
Wiley~\cite{Wiley2011} used Bayesian models created from sample
traffic to fingerprint obfuscated protocols (Dust~\cite{dust}, SSL and
obfs-openssh~\cite{obfs-openssh}) based on flow features, and found
that across these protocols length and timing detectors achieved
accuracy of 16\% and 89\% respectively over entire packet streams,
while the entropy detector was 94\% accurate using only the first
packet. A host's transport layer behavior (\eg the number of outgoing
connections) can be used for application classification.  Flow
properties can also be used to fingerprint the website a user is
visiting even if the flow is
encrypted~\cite{panchenko2011,Sun2002,Hintz2002,Bissias2005}.   

\mypara{Fingerprint Protocol Semantics (\fprSem)} A censor can
fingerprint flows based on protocol behaviour triggered through
different kinds of active manipulation, \ie by dropping, injecting,
modifying and delaying packets. The censor's goal is to leverage
knowledge of a protocol's semantic properties to elicit behaviour of a
known protocol. Alternatively, a censor can perform several
fingerprinting cycles to elicit the information on which to base
subsequent blocking decision.  In 2011, Wilde~\cite{Wilde2012}
investigated how China blocked Tor bridges and found that unpublished
Tor bridges are first scanned and then blocked by the Great Firewall
of China (GFW). Wilde's analysis showed that bridges were blocked in
the following fashion: \first When a Tor client within China connects
to a Tor bridge or relay, GFW's DPI box flags the flow as a potential
Tor flow, \second random Chinese IP addresses then connect to the
bridge and try to establish a Tor connection; if it succeeds, the
bridge IP/port combination is blocked.   

%% file: link-obfs.tex
\section{Link Circumvention} \label{sec:pt-model}

We now turn our attention from a censor's attack landscape to
censorship resistance systems. In Section~\ref{background}, we
mentioned that the link between information client and server
(\textsl{On Link} in \prettyref{fig:nw-info-circumvent}) is a frequent
target of censorship and hence the assumed threat model of most
resistance systems. It is common for resistance systems to employ an
intermediate proxy that divides the link into two parts, \first client
to proxy (\textsl{link circumvention}) and \second proxy to server.
The focus of this work is link circumvention: we first present an
abstract model of link circumvention, followed by an
\textsl{evaluation stack} that represents functional components of
link circumvention as a multi-layer stack. The evaluation stack can be
used as a common benchmark to visualize capabilities of different link
circumvention systems.

\input{pt-model}

\input{eval-stack}

We conduct a comprehensive survey of censorship resistance systems
that offer link circumvention (\pt). We identify \npapers such systems which
we broadly classify as ones that resist IP address/host blocking
(Section~\ref{ip-resist}), flow fingerprinting (Section~\ref{flow-resist})
and composite systems that combine the last two
(Section~\ref{comp-resist}). We further classify these high-level
categories where relevant. For each category, we discuss one
representative \pt in \textsl{depth} with respect to the \pt
evaluation stack (\prettyref{fig:eval1}) and the attack model
(\prettyref{fig:attack-tree}).  To capture the \textsl{breadth} of
systems in a given category, we also provide a brief discussion of
other relevant systems (under the title \textsl{Miscellaneous}). 

%% file: pt-model.tex
\subsection{Abstract Model of a Link Circumvention System}

We present an abstract model of a link circumvention system (\pt),
which lends itself well to analysis through the attack model outlined
in \prettyref{fig:attack-tree}. The goal of a \pt is to enable a
client application to communicate with a server application over the
Internet, even though direct connections are blocked
(\prettyref{fig:nw-info-circumvent}). The \pt-client exposes an API
whereby the client application can request that a communication
channel to the server application be opened. The \pt-client then
connects to the \pt-server over a blocking-resistant communication
channel, and the \pt-server connects to the server application. The
client application can then communicate with the \pt-client as if it
is communicating directly with the server.
 
The communication channel provided by the \pt-client and server has
similar properties to TCP.  Data sent through the channel will either
be delivered to the other end without corruption in the same order as
it was sent, or an error will be reported to the sender.  It is the
responsibility of the \pt to route communications between the
\pt-client and \pt-server, avoid blocking, and recover from any
corruption of data (whether by the censor or due to other network
disruption).

The \pt-client and \pt-server may be able to communicate directly over
the Internet, in that any intermediate networks or routers are not
aware of the protocol the \pt is using.  However in some cases there
may be a \pt-router which implements part of the \pt protocol so as to
facilitate the blocking resistant communication channel.

The \pt communication channel does not offer authenticity, so it is
the responsibility of the client and server applications to confirm
that data received on the channel originated from the expected party
and has not been corrupted in transit. However many practical \pts
will provide some degree of authenticity so as to meet the goal of
blocking resistance.

Ideally the latency of the communication channel will not be much
higher than that of the direct communication channel, but in some
cases a much higher latency is unavoidable in which case a client
application which is designed for a normal TCP connection may
malfunction.

This abstract model is implemented through the de-facto Pluggable
Transport standard~\cite{pt_spec_tor}. This API specifies how modules
are invoked (\ie as separate executable), how the communication
channel is implemented (\ie as an extension to the SOCKS protocol),
how configuration and status information is communicated between
client/server applications and Pluggable Transport client/server (\ie
through a combination of environment variables, command-line
parameters, and standard in/out/error descriptors).  Although this
specification was written for the Tor anonymity system, it is also
implemented in the Lantern~\cite{lantern} and Psiphon~\cite{Psiphon}
simple proxy CRSs.

%% file: eval-stack.tex
\subsection{Evaluation Stack} \label{sec:eval-stack}

\begin{figure*}%
\centering
\parbox{1.7in}{\includegraphics[width=1.6in]{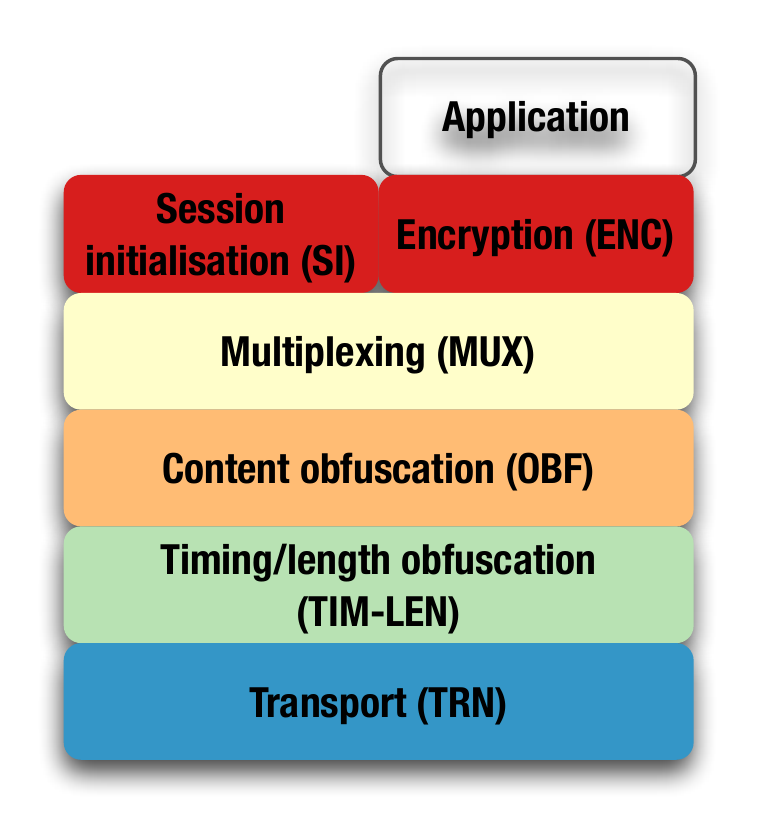}
\subcaption{\centering The abstract stack}}%
\qquad
\begin{minipage}{3.0in}%
\includegraphics[width=2.9in]{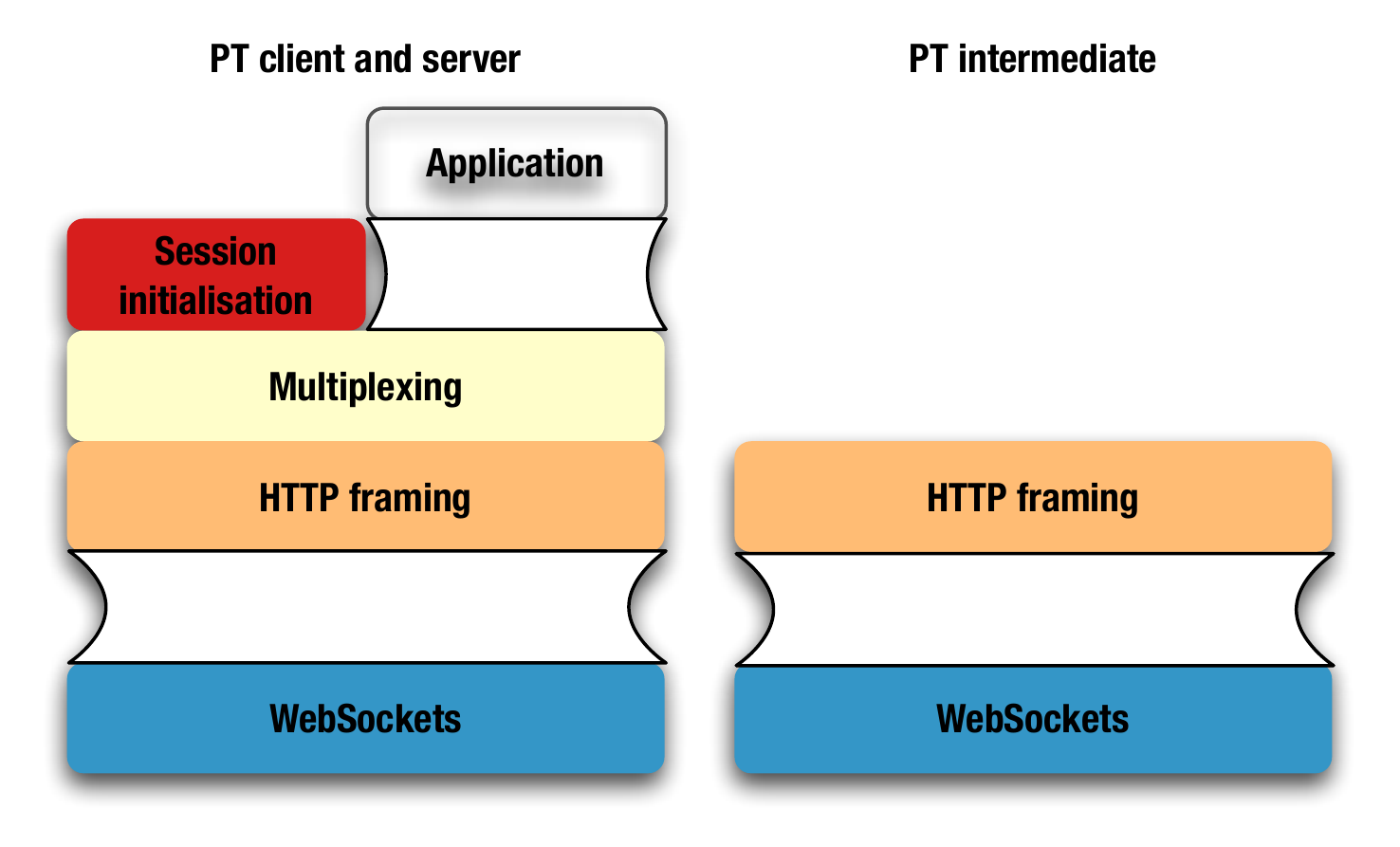}
\subcaption{\centering \flashproxy \label{fig:flashproxy}}
\end{minipage}%
\begin{minipage}{1.7in}%
\includegraphics[width=1.6in]{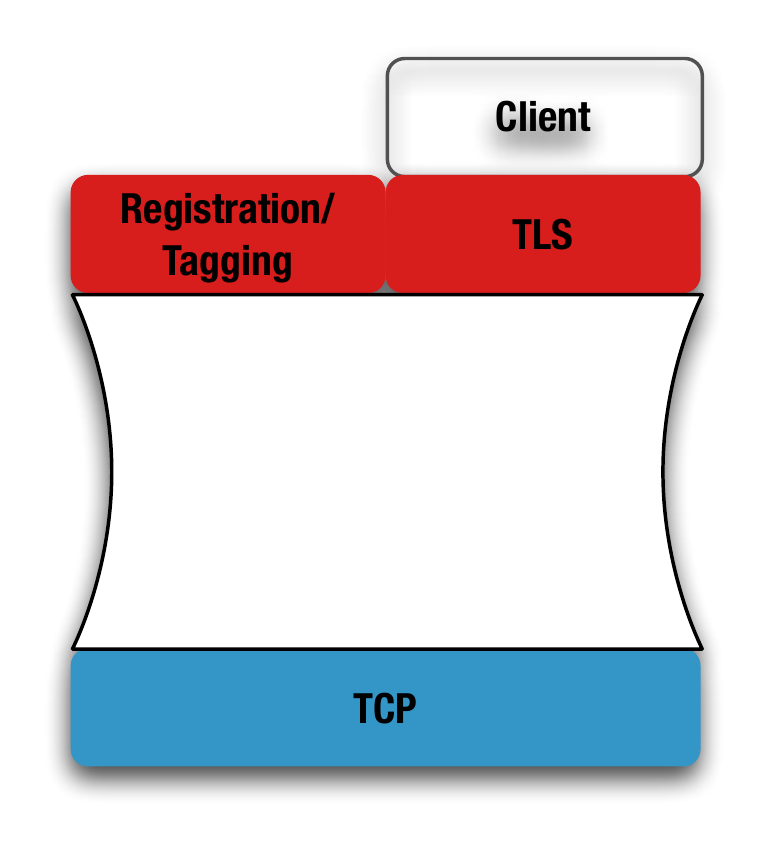}
\subcaption{\centering  \cirripede \label{fig:cirripede}}
\end{minipage}%
\captionsetup{justification=centering}
\caption{The Abstract Stack and Evaluation Stacks for \flashproxy and \cirripede.}%
\label{fig:eval1}%
\end{figure*}

In order to defend against the multiple avenues of attacks available
to a censor, a link circumvention system (\pt) is typically designed as a
series of components, with each component defending against one or
more attacks, either by itself or in conjunction with other
components. In order to describe the capability of each \pt, we
map each of them to a generic set of components shown in
\prettyref{fig:eval1}, arranged in layers analogous to a network
protocol stack.

The primary flow of payload information is between adjacent layers in
the stack, with the client/server application at the top and network
at the bottom. However just as with real-world network stacks, control
information does not always exactly follow this abstraction and may
skip layers. Also not all layers will be present in all \pts, as
some exclude certain attacks from their threat model.

Before data can be sent over the \pt communication channel, a session
must be established.  The \textbf{\session_init (\SI)} is responsible
for the handshake between the \pt-client and \pt-server. This may
involve negotiating connection parameters, performing authentication,
and deriving session keys.

On the same uppermost layer is the \textbf{\encrypt (\ENC)} which
takes application traffic data and encrypts it so as to look random to
an adversary.  The key for performing the encryption is provided by
the SI layer.

Next is the \textbf{\multiplex (\MUX)} layer, which allows multiple
application sessions to be multipexed over a single \pt channel, or a
single application session to be split over multiple \pt channels.
This layer is also responsible for error detection and re-assembly, if
the lower layers do not provide this.

Then the \textbf{\conObfs (\OBF)} formats the `random' data stream so
as to mimic the content of a different protocol, \eg HTTP or VoIP.

Next \textbf{\timeObfs and \lenObfs (\TIM-LEN)} hide the application's
timing and packet length patterns. The layer may perform the
obfuscation itself, or may just compute the changes which are
necessary and transmit this as control data to other layers which
actually delay and/or pad payload data.

Finally \textbf{\transport (\TRN)} is responsible for taking the
transformed payload data and sending it to the other side of the \pt
client/server pair.

%% file: ip-resist.tex
\section{IP Address/Host Filtering Resistance Systems}

\label{ip-resist}

A censor can disrupt access to a service by blocking the server's IP
address, or a key hop directly on path to the server (corresponding to
\blkRou in \prettyref{fig:attack-tree}). A number of tools have
emerged to resist IP address filtering which we further classify based
on the circumvention approach they take. Techniques under
\textsl{censorship surface augmentation} hide the censorship target
(\eg IP address) among a crowd such that censoring the target incurs
larger collateral damage than if it is blocked in isolation. In
\textsl{decoy routing}, a cooperative hop between client and server
applies circumvention-friendly treatment to packets containing a
special steganographic mark.

\subsection{Censorship Surface Augmentation} 

\label{attack-surface}

The accuracy of a censor's blocking decision depends on a number of
factors, such as the blocking mechanism employed and the quality of
target set to block. In particular, the censor's policy must make
consideration for the acceptable false positive rate as these have
political and economic ramifications~\cite{Tschantz2014}. Mechanisms
under this category obfuscate the censorship target in such a way that
censoring it incurs large collateral damage in terms of false
positives. 

\subsubsection{\flashproxy} \label{flashproxy}
\input{summaries/flashproxy}

\subsubsection{\others} \label{others-attack-surface}

\vpngate~\cite{vpngate:2014} is a public VPN service that randomizes
the IP addresses of its VPN servers through a pool of volunteer-run
\vpngate proxies. To prevent a censor from harvesting all proxies,
\vpngate \first only gives a fraction of the entire pool to each
client, \second includes decoy IP addresses (belonging to vitally
important hosts on the Internet, such as Windows Update servers) in
the pool, and \third by aggregating data across \vpngate proxies to
flag probes from a censor.  One design trend is to use a widely used
service as a proxy to fetch censored content. \meek~\cite{meek}
employs a technique called \textsl{domain
fronting}~\cite{domain_fronting:2015} to evade host-based censorship
by using an innocuous domain name (\textsl{front domain}) in the
unencrypted request header (\textit{TLS Server Name Indication header}
-- SNI), while hiding the domain of a proxy (\textsl{inside-domain})
in the encapsulated encrypted request (\textit{HTTP Host} header). The
front domain (the only one visible to a censor) is an intermediate web
service hosting many domains (typically a CDN) which decrypts the
inside-domain and internally routes the traffic to the relevant host
within its network. This host serves as a proxy for censored clients
to access blocked servers. \oss~\cite{oss} turns any existing Online
Scanning Service (OSS) (web services that take a URL as user input and
then fetch the web page behind that URL \eg PDFmyURL~\cite{pdfmyurl})
into a proxy to fetch censored content. A variation of this scheme is
for clients and servers to rendezvous on an intermediate host,
blocking which incurs significant collateral damage.
\cloudtransport~\cite{cloud-transport} clients and bridges (\ie
proxies) share a common account on a cloud storage which they use to
share files containing client requests and bridge responses in real
time. \collage~\cite{collage} peers exchange data through social
networking and photo sharing websites by embedding hidden messages in
user-generated content such as posts and images. \miab~\cite{miab}
improves \collage's rendezvous by leveraging \emph{blog pings}, \ie
real-time notifications a blog sends to a centralized network service
(a ping server) when content is updated.  By monitoring ping servers,
\miab peers automatically learn when a new message is available.
\defiance~\cite{defiance} allocates ephemeral IP addresses to its
gateways and bridges from a large pool of diverse IP addresses.  The
transience and diversity of IP addresses make it hard for a censor to
block or enumerate them. To access blocked websites, the \defiance client
must connect to a shorted-lived bridge which acts as a proxy.  To
learn an ephemeral bridge location, a client must successfully
complete a \textit{dance}; that is, make a sequence of pre-agreed
timed short-lived connections to ephemeral gateways.

\subsection{Decoy Routing}

\label{decoy-routing}

In this approach, clients covertly signal a cooperating intermediate
router to deflect their traffic meant for a non-blocked destination
(to evade the censor) to a blocked one. To deflect traffic, deflecting
routers must be located on the forward network path from the client to
the non-blocked destination.  These routers must therefore be
strategically positioned to optimise the number of censored users that
can be served~\cite{optimizing_proxy_placement}.  It is theoretically
possible for a censor to defeat decoy routing by routing traffic
around the deflecting routers~\cite{routing_around_decoys}. However,
in practice this is believed to be too costly for a censor because of
business relationships with other ISPs, and monetary, performance and
quality of service degradation issues thereby
induced~\cite{decoy_routing_costs}.

\subsubsection{Cirripede} \label{cirripede}
\input{summaries/cirripede}

\subsubsection{\others} \label{other-decoy}

\telex~\cite{telex}, \tapdance~\cite{tapdance} and
\curveball~\cite{foci11-decoy,curveball} can selectively \textit{tag}
individual connections on-the-fly (in contrast to \cirripede that
deflects client traffic \textsl{after} registration). \telex and
\curveball embed their tag in the random nonce of a TLS handshake with
a non-blocked destination. \tapdance encodes the tag in a connection's
incomplete HTTPS request to a decoy server using a novel steganography
scheme. \cirripede, \tapdance and \curveball support asymmetric flows
where upstream and downstream traffic do not follow the same path
because of ISP's internal routing.  \tapdance is the only solution
that does not require active inline flow blocking to prevent further
decoy-server client communication. This property makes it easier to be
deployed by ISPs without disturbing existing traffic and quality of
service.  \id-based-crypto-tagging~\cite{id-based-crypto-tagging}
improves decoy routing solutions in general by simplifying key
distribution and providing forward secrecy. These are achieved with
the use of identity-based encryption instead of traditional public-key
cryptography.

%% file: summaries/flashproxy.tex
Simple proxies with long-lived IP addresses that circumvent IP
blocking by relaying traffic between a client and censored server
suffer from the problem of being enumerated and subsequently blocked
by a censor (\fprRou, \corRou and \blkRou nodes in the attack diagram
in~\prettyref{fig:attack-tree}). 

To protect against these attacks, \flashproxy introduces two new
entities, \ie \textit{facilitators} and \textit{flashproxies}
(\prettyref{fig:flashproxy}, left and right, respectively).  A
\emph{facilitator} is a volunteer website outside the censored region
which may be blocked by the censor, yet remains reachable through
low-bandwidth channels such as email. Over this channel, the
\session_init module (\SI) of a censored user registers itself by
sending its IP address and a port where it awaits incoming
connections.  On the \textit{facilitator} side, the \SI adds a special
\textit{badge} in all the web pages it serves to uncensored users,
typically a piece of Javascript. When an uncensored user visits a web
page on the \textit{facilitator} website, the \textit{badge} turns the visitor's
web browser into a \textit{flashproxy}.  The \SI of a
\textit{flashproxy} (running in the web browser of an uncensored user)
connects to the \textit{facilitator} to retrieve an IP-port pair for a
censored user and initiates a connection to it. The \SI of the
censored user accepts the connection to complete the rendezvous
between the two entities.  Thereafter, the \conObfs (\OBF) of the
\textit{flashproxy} relays censored content for the censored user
through HTTP (``HTTP framing'' in Figure~\ref{fig:flashproxy}).

\flashproxy offers resistance against IP address blocking only,
consequently leaving a number of paths on the attack diagram exposed.
A censor could block traffic based on content (\corCon), use
statistical traffic properties (\fprLen and \fprTim) to detect the
protocol or content to block, or observe characteristic patterns in
incoming connections to censored hosts (\fprSem). The authors suggest
using \flashproxy in combination with Tor to thwart \corCon attacks.

%% file: summaries/cirripede.tex
Cirripede offers resistance against IP address filtering (\blkRou,
\fprRou and \corRou in the attack diagram
in~\prettyref{fig:attack-tree}). In addition to this, it also resists
content-based fingerprinting (\fprCon) and content tampering/blocking
(\corCon) through its authenticated encryption layer.  Its
corresponding evaluation stack is presented in
\prettyref{fig:cirripede}. 
 
To use \cirripede, a censored user must first register its IP address
and a shared secret with a Registration Server (RS). The registration
is facilitated by a friendly ISP that deploys Deflecting Routers (DRs)
that redirect non-registered client traffic to the Registration
Server. To register, the \session_init (\SI), \transport (\TRN) and
\encrypt (\ENC) module of a client coordinate to encode a covert
registration signal into the \textit{TCP Initial Sequence Number
(ISN)} of a series of packets destined to a non-blocked destination.
The registration packets are deflected by the Deflecting Routers (DRs)
and reach the Registration Server where they are inspected. If the
Registration Server successfully recognises the signal in the packets,
its \SI instructs all Deflecting Routers within the ISP network to
deflect subsequent client traffic to a Service Proxy (SP). After
sending the registration packets, a client selects an innocuous
non-blocked destination and initiates a TLS handshake to it which is
taken over by the Service Proxy that acts as a web proxy to censored
content thereafter.

A censor could still attack unprotected nodes of the attack
diagram in Figure~\ref{fig:attack-tree}. Traffic from/to different websites
generally has different characteristic patterns, therefore a censor
could determine that traffic feigned to originate from a non-blocked
website actually comes from a blocked website (\fprLen and \fprTim).
It could also detect protocol implementation inconsistencies due to
the non-blocked and blocked destination running different software
stacks (\fprSem). 
%\cirripede is also vulnerable to replay attacks: a censor who replays
%the registration packets will find itself incapable of decrypting the
%TLS traffic after the handshake: this hints that a new key is
%covertly negotiated during the TLS handshake.

%% file: flow-resist.tex
\begin{figure*}%
\centering
\parbox{1.6in}{
\includegraphics[width=1.5in]{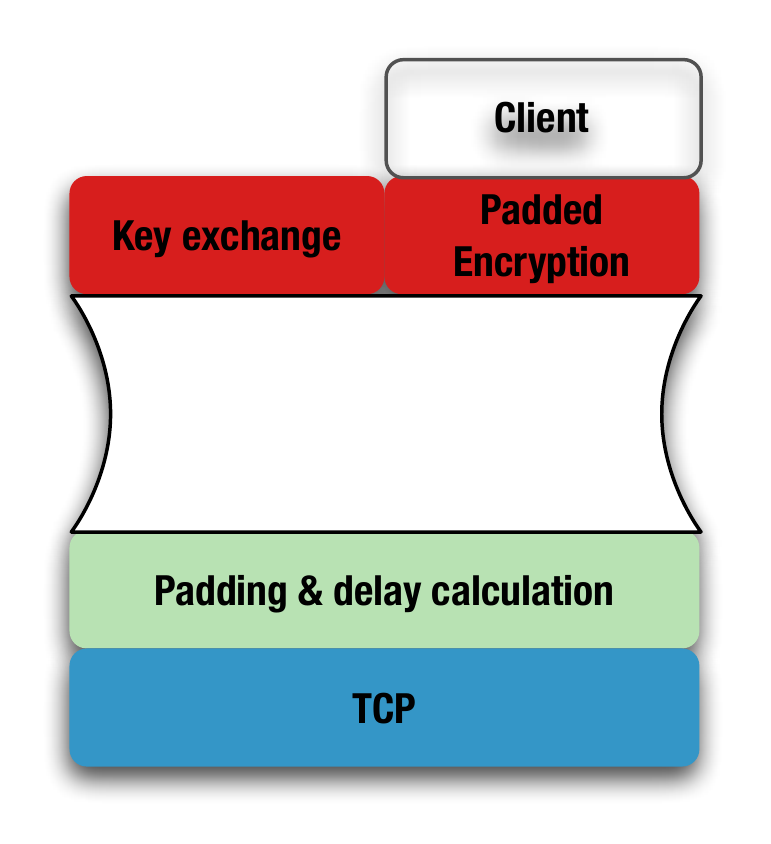}
\subcaption{\centering \scramblesuit \label{fig:scramblesuit}}}%
\qquad
\begin{minipage}{1.6in}%
\includegraphics[width=1.5in]{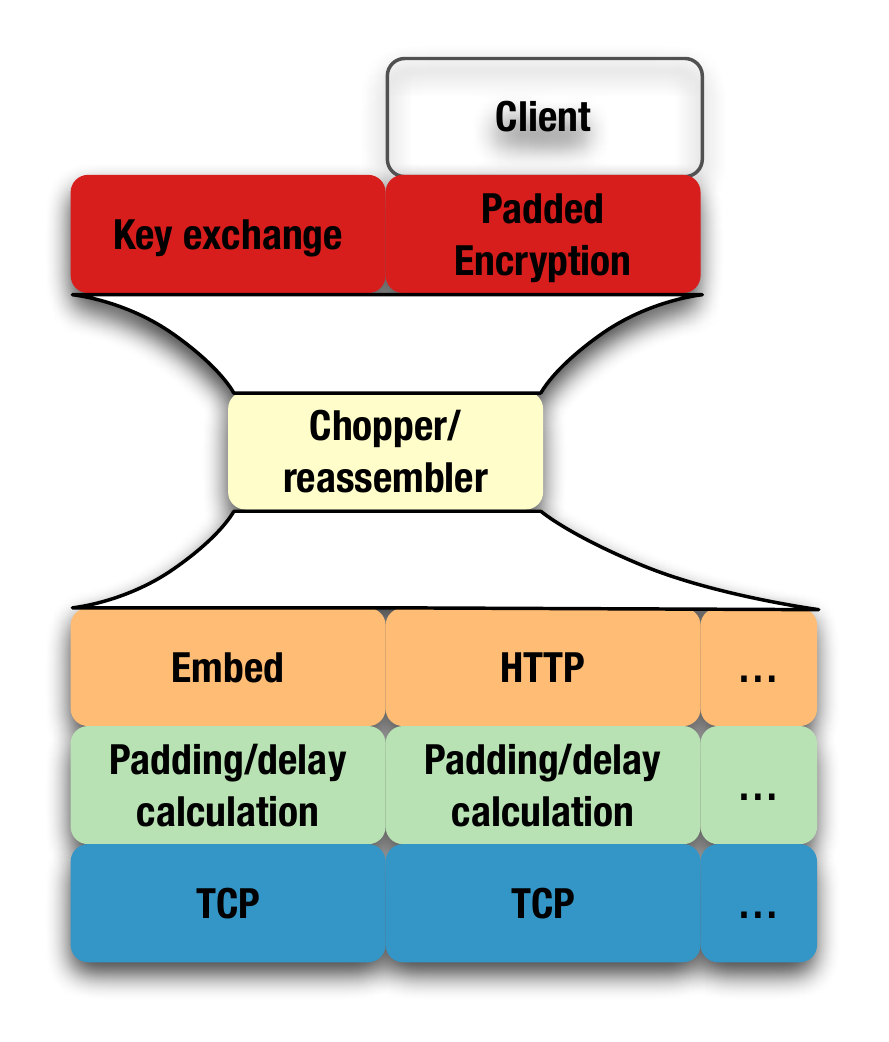}
\subcaption{\centering \stegotorus \label{fig:stegotorus}}
\end{minipage}%
\begin{minipage}{1.6in}%
\includegraphics[width=1.5in]{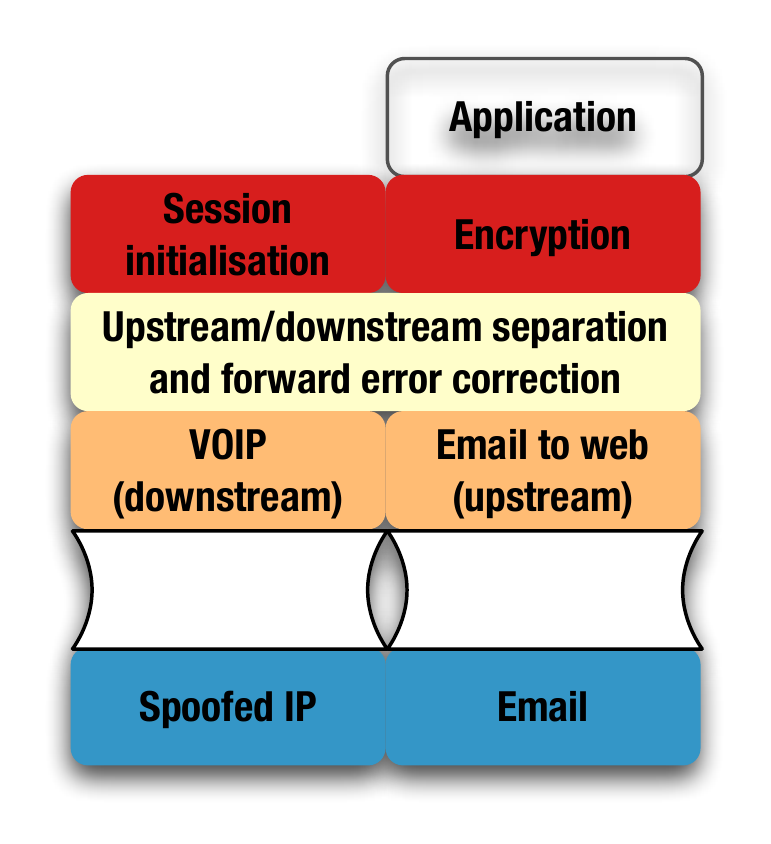}
\subcaption{\centering \censorspoofer \label{fig:censorspoofer}}
\end{minipage}%
\begin{minipage}{1.6in}%
\includegraphics[width=1.5in]{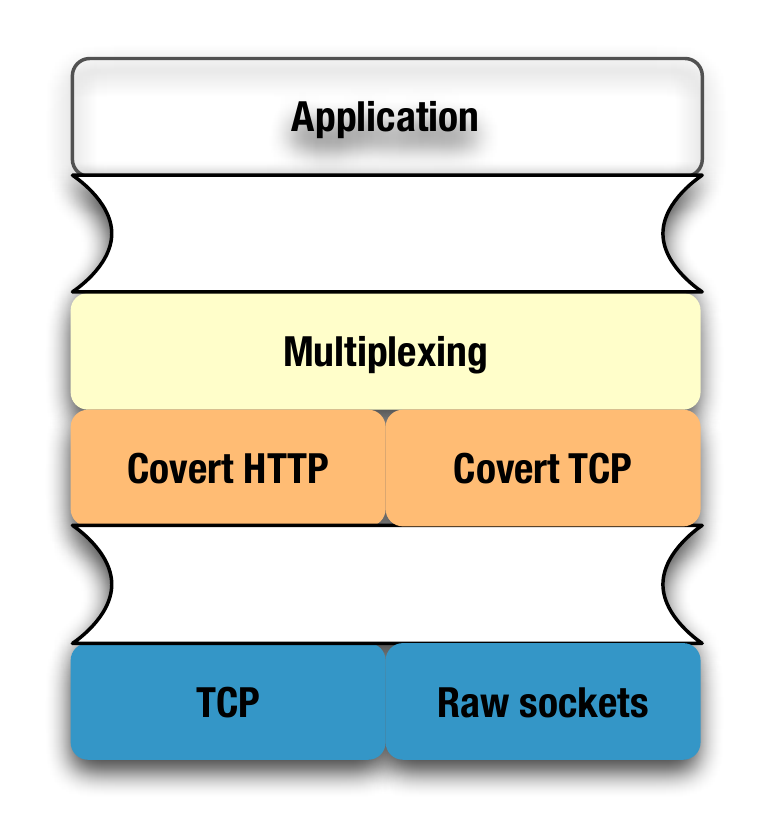}
\subcaption{\centering Khattak \detal~\cite{Khattak2013} \label{fig:khattak}}
\end{minipage}%
\captionsetup{justification=centering}
\caption{Evaluation Stacks for \scramblesuit, \stegotorus, \censorspoofer and Khattak \detal~\cite{Khattak2013}.}%
\label{fig:1figs}%
\end{figure*}

\section{Flow Fingerprinting Resistance Systems} 

\label{flow-resist}

These systems obfuscate blocked traffic such that it cannot be
fingerprinted by a censor, some by offering resistance against
\textsl{fingerprinting of protocol semantics}, while others
\textsl{mimic} a supposedly whitelisted category. Another approach to
censorship evasion is to transform flows leveraging a censor's
analysis limitations (\textsl{monitor-driven flow transformation}).

\subsection{Fingerprinting of Protocol Semantics} 

\label{protocol-fingerprint}

A number of schemes offer resistance against protocol
scanning, \ie techniques used by a censor to confirm that a machine is
indeed part of an anti-censorship system. Typically, a censor probes
for an open port and attempts to ``speak'' the anti-censorship
protocol. To defeat protocol scanning, a
\silentknock~\cite{silentknock} server accepts incoming connections on
a particular port only from clients that authenticate with a special
``knock'' (a one-way authentication mechanism embedded in TCP
headers). \bridgespa~\cite{bridgespa} (originally known as
\spator~\cite{spator}) builds upon \silentknock's design but relaxes
server-side memory constraints associated with per-client
housekeeping. It replaces counters with rounded-to-the-minute UTC
timestamps and long-lived keys with short-lived ones.
\defiance~\cite{defiance} imposes several levels of address
indirection to prevent unauthenticated access to bridges from protocol
scanning censors.
\keyspacehopping~\cite{keyspace_hopping:2003} discloses 
only a fraction of its proxies to each client.
It defeats protocol scanning probes through the use of shared secrets. It
forces a censor to dedicate more resources to discover proxies by 
requesting clients to solve computational problems.

\subsection{Mimicry} \label{mimicry} Mimicry-based mechanisms
transform traffic to look like whitelisted communication, such that
the transformed traffic resembles the syntax or content of an allowed
protocol or randomness.  

\subsubsection{Mimic Existing Protocol or Content} 

A large body of work evades censorship by imitating an innocuous
protocol (\eg HTTP) or content (\eg HTML). Typically the mimicry is
based on widely-deployed protocols or popular content thus
complicating a censor's task by \first increasing the censor's workload
as there is more volume of traffic to inspect, and \second
increasing the collateral damage associated with wholesale protocol
blocking.

\mypara{\stegotorus}
\input{summaries/stegotorus}

\mypara{\others}
\foe~\cite{foe} and \mailmyweb~\cite{mailmyweb} proxies
send users static web pages as email attachments in
response to censored URLs they receive in an email \textsl{Subject}
field (assuming that SMTP is not monitored or email traffic is
encrypted). \sweet~\cite{sweet} creates a bi-directional communication
channel by encapsulating censored traffic into email attachments such
as images.  To prevent a censor from blocking all emails sent to the
proxy, each client sends requests to a unique email address.
\skypemorph~\cite{skypemorph} shapes inter-packet timing and packet
size distribution so as to mimic a Skype video call.
\transteg~\cite{transteg} negotiates an \emph{overt codec} during a
VoIP call initialisation, but thereafter encodes the raw voice stream
with a different lower-bitrate codec (the \emph{covert codec}), thus
reducing the length of packets to transmit. The freed space is filled
with low-bandwidth covert traffic. \fte~\cite{fte} extends
conventional symmetric encryption with the ability to specify the
format of the ciphertext with a regex. This thwarts protocol
fingerprinting by transforming a blocked \emph{source}
application-layer protocol into an unblocked \emph{target}
application-layer protocol.

Protocol imitation has a number of limitations that make it possible
to distinguish imitated traffic from legitimate
traffic~\cite{parrot,Geddes2013}. To avoid the above pitfalls, a
number of recent mimicry-based circumvention systems reuse genuine
software and libraries and tunnel covert traffic through them.
\facet~\cite{facet} streams potentially censored videos over
video-conferencing calls of popular applications such as Skype, Google
Hangout and FaceTime. \freewave~\cite{freewave} modulates traffic into
acoustic signals and streams them directly into an existing VoIP
application such as Skype. \jumpbox~\cite{jumpbox:2014} feeds HTTP
request directly into a web browser's stack in order to `emulate' the
networking characteristics of the HTTP protocol.
\castle~\cite{castle_game:2015} provides obfuscation for low-bandwidth
application (\eg chat and emails) through real-time strategy (RTS)
video games. It encodes data as player actions in the game and uses
desktop-automation software to feed player actions directly to the
game's user interface. \rook~\cite{rook_video:2015} leverages First
Person Shooter (FPS) video games to obfuscate chat sessions of users
within a censor region. It identifies packet fields of high entropy
and replaces them with other legitimate game values before they are
sent.  \marionette~\cite{marionette:2015} mimics a protocol's stateful
semantics and statistical properties by composing probabilistic state
machine automata through flexible plugins and domain-specific
languages. Content obfuscation is achieved through template grammars
built using probabilistic context-free grammars.

\subsubsection{Mimic Unknown Protocol or Content} 

\label{mimic-random}

Another approach is to make traffic look like an
unknown protocol, either by imitating randomness or arbitrarily
deviating from a known blocked protocol. This idea is motivated by the
general assumption that a censor implements blacklisting of known
protocols and is unwilling to incur high collateral damage associated
with whitelisting. 

\mypara{\scramblesuit}
\input{summaries/scramblesuit}

\mypara{\others} \mesg-stream-encrypt~\cite{mesg-stream-encrypt} is
the de-facto obfuscation mechanism used by BitTorrent. Both the key
exchange and traffic session contain only random-looking bytes.
Padding is added to all traffic including the handshake.
\dust~\cite{dust} is a similar system with the capability to shape
packet sizes based on an arbitrary distribution.
\obfsTwo~\cite{obfs2} obfuscates traffic content with encryption, but
the key exchange does not provide authentication against passive and
active attackers.  \obfsThree~\cite{obfs3}, initially adopted by
\scramblesuit, improves \obfsTwo by negotiating keys using anonymous
Diffie Hellman (DH) with a special encoding so as to be
indistinguishable from a random string.  This forces a censor to
actively probe the server or perform a man-in-the-middle attack to
detect the key exchange. \obfsFour~\cite{obfs4} adds authentication to
\obfsThree using the ntor
handshake~\cite{ntor_academic_paper:2013,ntor_specs} and is now used
by \scramblesuit. Unlike \scramblesuit, \obfsTwo and \obfsThree do not
randomise packet lengths and inter-packet timings.

\subsection{Monitor-driven Flow Transformation (MDFT)}
\label{monitor-driven}

Systems under this category shape traffic in a way
that exploits the limitations of an adversary's traffic fingerprinting
model. Effectively, this mechanism can provide unobservability to even
cleartext traffic, which is particularly useful for countries that
prohibit encrypted traffic.    

\subsubsection{Khattak \detal~\cite{Khattak2013}} \label{khattak}
\input{summaries/khattak}

\subsubsection{\others} \label{other-monitor} \gohop~\cite{gohop:2014}
breaks the notion of `flow' used by Deep Packet Inspection (DPI) boxes
by sending traffic over a set of randomized ports. Clayton
\detal~\cite{clayton:pet2006} note that the GFW terminates offending
connections through injection of TCP reset packets. A client can
circumvent this form of censorship by ignoring all the TCP reset
packets it receives.  West
Chamber~\cite{westChamberS1,westChamberS2,westChamberS3} forces the GFW to
purge state for a connection and exclude it from censorship analysis
by exchanging specially crafted TCP reset packets between the client
and server. These packets are ignored by TCP stacks of the client and
server, but considered by the GFW.

%% file: summaries/stegotorus.tex
\stegotorus obfuscates packet length and inter-packet timing of Tor
traffic thus thwarting fingerprinting of flow properties (\fprLen and 
\fprTim nodes in the attack diagram
in~\prettyref{fig:attack-tree}).  It also prevents content
fingerprinting (\fprCon) and content tampering/blocking (\corCon) with
authenticated encryption. Optionally it can mimic a set of innocuous
protocols over which covert traffic is sent.  The corresponding
evaluation stack is presented in \prettyref{fig:stegotorus}.

\stegotorus comprises two modules, both of which can be implemented by
a combination of \pt components.  To start a session, the \SI of a
client and server (\ie proxy) first establish a shared key through a
key-exchange that only contains random bytes: this thwarts
content-based filtering (\fprCon). Once a session is established, the
\TIM-LEN module chops fixed-length input packets into random-sized
messages. Sizes are taken from a trace of an innocuous protocol
session pre-recorded by the user or bundled with the software
(\fprLen). A 32-byte ID is added to messages so they can be re-ordered
by the recipient.  The \ENC module then encrypts messages individually
(\corCon and \fprCon) and passes them again to the \TIM-LEN module
which adjusts their sending time in accordance with packet
inter-arrival times derived from a pre-recorded traffic trace
(\fprTim). Optionally, the \conObfs (\OBF) component of \stegotorus
can mimic a known protocol such as unencrypted HTTP. For example the
\OBF of a client may embed the encrypted messages into the HTTP
\textit{Cookie} header and url part of requests, and the \OBF of a
server may reply by steganographically embedding content into the body
of HTTP responses.

\stegotorus uses long-lived server locations. So if a censor manages
to harvest them, it could block them by injecting TCP \textit{RST}
packets or erroneous DNS responses (\corRou, \blkRou and \fprRou).

%% file: summaries/scramblesuit.tex
\scramblesuit~\cite{scramblesuit} obfuscates a censored protocol with
random-looking bytes for all its traffic including session
initialisation thus thwarting fingerprinting based on flow content
(\fprCon in the attack diagram in~\prettyref{fig:attack-tree}).
Encryption of content provides content obfuscation (\fprCon),
confidentiality and resistance against tampering (\corCon). Packet
lengths and inter-arrival times are also randomised (\fprLen and
\fprTim). \scramblesuit uses a special protocol for session
bootstrapping that resists tampering (\corCon) and protocol scanning
(\fprSem). The corresponding evaluation stack is presented in
\prettyref{fig:scramblesuit}.

To connect to a \scramblesuit proxy, the \SI of a client redeems a
short-lived ticket retrieved from a low-bandwidth out-of-band channel.
After the first authentication, the server gives the client a ticket
for the next connection. This ticket provides mutual authentication
and therefore resistance against protocol scanning from a censor
(\fprSem). Furthermore, it only contains random bytes to evade
content-based detection (\fprCon). Once a session is initialised, the
\ENC component turns all traffic into random-looking bytes, thereby
thwarting content-based fingerprinting and blocking (\fprCon). It also
provides authentication and confidentiality (\corCon).  The \TIM-LEN
component randomises packet lengths (\fprLen) and timing of flows
(\fprTim) using discrete probability distributions provided by the
\ENC module.

\scramblesuit protects against most attacks in the attack diagram
(Figure~\ref{fig:attack-tree}). One limitation is that the IP
addresses of \scramblesuit proxies are known, and therefore vulnerable
to being enumerated and subsequently blocked by a server (\corRou and
\blkRou).

%% file: summaries/khattak.tex
This work explores protection against content-based filtering (\fprCon
and \corCon in the attack diagram in~\prettyref{fig:attack-tree}) by
exploiting limitations of the implementation of Deep Packet Inspection
(DPI) boxes.  The evaluation stack is presented in
\prettyref{fig:khattak}.

Being operationally similar to Network Intrusion Detection Systems
(NIDS), DPI boxes grapple with the same issues that make them
vulnerable to evasion: when to create state for a flow to monitor,
when to tear down state for a flow being monitored, how to interpret a
packet stream in the presence of multiple routing paths, divergent
header fields, and overlapping IP fragments and TCP segments. The
authors intermingle a number of specially crafted packets in their
regular packet stream to fetch Web content from a server outside China
using a client inside China, and highlight a number of vulnerabilities
in the Great Firewall of China (GFW)'s traffic analysis model. For
example, one observation was that by sending a TCP reset packet with
low \textit{TTL} value for an existing connection, subsequent packets
containing blacklisted keywords are no longer censored. 

Most nodes of the attack diagram (Figure~\ref{fig:attack-tree}) are
not protected by this approach.  Even without a clear view of the
entire stream, a censor could still block traffic on a per-packet
basis, for example by filtering packets destined to known server IP
addresses (\fprRou, \corRou and \blkRou) or by fingerprinting packet
length (\fprLen).

%% file: comp-resist.tex
\section{Composite Censorship Resistance Systems}

\label{comp-resist}

These systems offer resistance against filtering of both IP addresses
and hosts (Section~\ref{ip-resist}), and flow-based censorship (Section~\ref{flow-resist}).   

\subsection{\censorspoofer}\

\label{censorspoofer} 

\input{summaries/censorspoofer}

\subsection{\others}

\label{other-composite}

\freewave~\cite{freewave} mimics VoIP traffic while also hiding proxy
IP addresses. It modulates traffic into acoustic signals and sends
them over a VoIP network such as Skype, Vonage or iCal. To hide the IP
address of a proxy, client traffic is relayed through multiple VoIP
peers. For Skype, this is achieved by configuring the \freewave proxy
as an ordinary node so that VoIP traffic is routed via Skype
\textit{super nodes}. \infranet~\cite{infranet} thwarts IP blocking by
turning a non-blocked cooperative website into a proxy.  Besides
relaying censored content, proxy websites continue to serve their
usual uncensored content. For its upstream channel, \infranet encodes
covert traffic in sequences of HTTP requests; for its downstream
channel, it steganographically embeds content in uncensored images.

%% file: summaries/censorspoofer.tex
\censorspoofer provides resistance against IP address harvesting
(\blkRou, \corRou and \fprRou nodes in the attack diagram
in~\prettyref{fig:attack-tree}) using spoofed source IP address, and
resistance against content-based blocking (\fprCon and \corCon) by
mimicking encrypted VoIP traffic. The evaluation stack is presented in
\prettyref{fig:censorspoofer}.

\censorspoofer client sends requests for censored content to a
\censorspoofer server (\textit{spoofer}) over email, which then serves
blocked content to the client over a VoIP call. The \SI of a
\censorspoofer client creates four accounts: two email accounts (one
from a local provider and one from a foreign one), and two VoIP
accounts (one from a local registrar and one from a foreign one). The
local accounts serve as client agents, while the foreign accounts act
as server agents. A client registers with the \textit{spoofer} by
sending a registration message containing information related to its
accounts and a shared cryptographic key. The \textit{spoofer} logs in
to the foreign accounts (VoIP and email) and monitors incoming
messages from the client's local accounts. Once registered, the \SI of
a client initiates a session by calling the \textit{spoofer}'s VoIP
account.  The \textit{spoofer} accepts the call and provides a dummy
IP address and port. The client then sends dummy UDP traffic to the IP
address provided by the \textsl{spoofer}, while sending censored
requests to the \textsl{spoofer}'s email account.  Upon receiving a
request, \textit{spoofer} serves the blocked content in encrypted UDP
traffic by spoofing source IP address of the dummy host. The \ENC
module of \censorspoofer prevents content-based censorship (\fprCon)
and provides data confidentiality and authentication (\corCon).

As \censorspoofer does not shape traffic, a censor may be able to
detect discrepancies between a real voice call and the web traffic
sent over UDP (\fprLen and \fprTim in Figure~\ref{fig:attack-tree}).
Furthermore, there may be an indicative correlation between the timing
of a VoIP call and SMTP traffic (\fprSem).

%% file: discussion.tex
\input{tables/tb_summary.tex}

\section{Discussion and Challenges}

\label{sec:discussion}

In \prettyref{tab:sumpt}, we provide a summary of link circumvention
systems (\pts). For each \pt, we note components of the evaluation
stack (\prettyref{sec:eval-stack}) that the system implements to offer
protection along various path(s) in the attack diagram
(\prettyref{fig:attack-tree}). We note that a large number of systems
resist content-based blocking and fingerprinting (\fprCon and
\corCon), with relatively less focus on routing-based censorship
(\blkRou, \corRou and \fprRou). Only a few systems protect against
flow fingerprinting based on length and timing (\fprLen and \fprTim),
while \textsl{none} address attacks related to manipulation of
protocol semantics (\corSem). The heavy emphasis of existing \pts on
content-based circumvention is in contrast to how censors have
historically implemented blocking in practice, \ie through blocking or
corruption of routing information~\cite{sadia2015hotpets}.
Furthermore, \pts tend to cluster around specific evasion styles
making them vulnerable to unprotected paths in the attack diagram
(Figure~\ref{fig:attack-tree}).  

Our systemization highlights two main limitations in the area of link
circumvention, \ie lack of \first common evaluation criteria, and
\second modularity and flexibility. Currently each \pt uses its own
evaluation criteria which makes it hard to assess the scope of
circumvention offered by a system in isolation as well as in
comparison with other systems. Testing and benchmarking \pts is
complicated due to the absence of a suitable testbed, and involves
separately downloading, compiling and using systems with potentially
different (and conflicting) build and run environments. This has led
to a situation where research and development efforts are concentrated
on proposing new schemes instead of improvising existing systems.
Related to the testbed issue, there is no environment to simulate
adversaries. Consequently, most adversarial assumptions are
theoretical and in some cases disconnected from how real censors
operate~\cite{sadia2015hotpets}. A number of systems protect against a
censor capable of fingerprinting flow properties and content.
Evaluation of such systems will greatly benefit from a repository of
traffic capture files. Systems that perform traffic shaping use
various protocols in a `correct' manner and require labelled datasets
against which to validate their schemes. Adversary
Lab~\cite{AdversaryLab} has done some preliminary work on developing a
standard environment to evaluate systems resistant to flow
fingerprinting by subjecting them to a range of adversaries.  Another
important but overlooked aspect of evaluation is performance and
usability. This is particularly important because \pts often rely on
help from volunteers for deployment who are naturally interested in
the system's performance and maintenance costs. We note lack of
principled performance evaluation, with some metrics being misaligned
to a given use case.  For example, some systems use file download to
approximate web browsing performance.  There has been preliminary work
recently on conducting user studies to evaluate \pt's
performance~\cite{david2015hotpets}.

The ability to quickly develop link circumvention systems (\pts) for
particular locations (spatial agility) and in response to changes
(temporal agility) is particularly important for censorship resistance
because there is no one approach which is optimally efficient and
resistant to all attackers.  However, building a new \pt is a
significant undertaking, in particular if it must include several
classes of blocking-prevention techniques (\eg scanning-resistance,
traffic-analysis resistance, encryption) and robustness techniques
(e.g. flow control, error detection and recovery). In many situations
having several schemes in operation is more effective than using any
one~\cite{ElahiMG16} (for example, Tor dealt with TLS handshake
fingerprinting of January and September 2011 by modifying its protocol
to protect against the vulnerable attack path).  Recently, there has
been a case for combining multiple \pts (particularly in the context
of Tor's Pluggable Transports which are essentially an instantiation
of our abstract model of link circumvention systems) with the goal to
offer circumvention tailored to different censorship scenarios. So
far, the sharing of Pluggable Transport features has happened not in a
black-box way, but through the sharing of source code. LibFTE is in
use by Tor (in its fteproxy Pluggable Transport form) and a number of
other projects.  Similarly, \meek which was originally developed for
Tor now also exists in a fork by Psiphon~\cite{Psiphon} with minor
adaptations.  Fog~\cite{Fog} uses multiple proxies to chain Pluggable
Transports in a black box fashion.  This approach is not suitable for
practical deployment due to a number of limitations.  For example, not
all combinations of Pluggable Transports make sense: the chain
\obfsThree (flow fingerprinting resistance) followed by \flashproxy
(IP address filtering resistance) offers more comprehensive
resistance, but the reverse, i.e. \flashproxy followed by \obfsThree
breaks the former's network layer assumptions. 

The evaluation stack described in \prettyref{sec:eval-stack} was
designed to help understand and evaluate link circumvention schemes
(\pts), as well as assist the rapid development of new
schemes. We propose a framework to build link
circumvention schemes out of reusable components following the
evaluation stack architecture which we call \emph{Tweakable
Transports} (we are currently writing the design
specification~\cite{tweakable}).  

Tweakable Transports provide a modular and flexible platform for
writing link circumvention systems (\pts), while simplifying
evaluation.  Following the evaluation stack, components from a link
obfuscation scheme can be extracted so that each component complies
with the abstract model the stack defines. This approach assists the
design process by providing a set of patterns to follow, and a
methodology for evaluating the censorship resistance features which
are offered.  Just as abstractions for components have been developed
for compiler design (lexer, parser, code generator) or GUI design
(model, view, controller), a systematic approach to design reduces
development time and improves quality of code. A particular instance
of a Tweakable stack may be designed by an expert familiar with the
properties of each component and a censor's blocking techniques and so
allow the trade-off between performance and censorship resistance.
Alternatively instances could be automatically generated and tested
against the real censorship system or a simulation of one, so as to
quickly find an adequate link circumvention scheme.

Tweakable Transports facilitate combining different link circumvention
(\pt) features.  Each component can be replaced with another which is
compatible and components can be inserted or removed.  This approach
allows code-reuse because a component developed for one Tweakable
Transport can be used for another. In so doing, more collaboration
opportunities are allowed, better testing can be performed on
frequently required components improving reliability and both spatial
and temporal agility. As a result Tweakable Transports exponentially
increases the number of possible link obfuscation scheme. The
development effort to add one component adds not just one new link
obfuscation scheme, but creates a whole new family of schemes, each
one of which the censor will need to test against any proposed
fingerprinting or blocking technique. The increased development cost
and possibility of false-positives reduces the likelihood that a
censor will be able to effectively block the resulting \pts.

From the link obfuscation schemes summarised in this paper, adapting
their implementations to be Tweakable Transports allows weaknesses to
be addressed.  Missing layers (\eg resistance to timing and
packet-length fingerprinting) leave some schemes open to attack.
Rather than developing a component from scratch, a component can be
imported from another link obfuscation scheme. New schemes can also be
created, such as combining the content obfuscation and timing/length
obfuscation with a common session initialisation and encryption
component, via a multiplexing component.

One way to build link obfsucation scheme designs following the
principles of Tweakable Transports is to create an common API
specification that allows components to
communication~\cite{tweakable}. This approach has the advantage of
allowing existing code to be re-used, which is particularly important
when protocols to be impersonated are not specified and only available
as binary modules.  However components need to be modified to fit the
API specification. An alternative approach taken to creating Tweakable
Transports is taken by Marionette where the full link obfuscation
scheme is written in a domain specific language (DSL) optimised for
this case~\cite{marionette:2015}, reducing flexibility but also
reducing effort for cases when the DSL is sufficient.

%% file: tables/tb_summary.tex
\begin{table*}[!htbp] 
\centering

\caption{A summary of link circumvention systems. Columns represent a node of the attack diagram (\prettyref{fig:attack-tree}). Within each column, different symbols denote components of evaluation stack (\prettyref{fig:eval1}) that a system implements to protect against the attack represented by the column. Symbols map to the evaluation stack as follows: \si{}\session_init (\SI), \enc{}\encrypt{} (ENC), \mux{}\multiplex{} (MUX), \cobf{}\contentObfs (OBF), \tobf{}\timeObfs (TIM-LEN), \lobf{}\lenObfs{} (TIM-LEN) and \trn{}\transport (TRN).}

\resizebox{\textwidth}{!}{%
%\scalebox{1} {
%\hspace*{-3.6cm}

{\setlength{\extrarowheight}{5pt}%
\begin{tabular}[12pt]{ c | lc | c |  c |  c |  c ? c | c | c | c | c }
\label{tab:sumpt}\\

 & & \multirow{1}{*}{Section \#} & \multicolumn{4}{c?}{Blocking} & \multicolumn{5}{c}{Fingerprinting} \\

\cline{2-12}

 &  &   &  \corCon & \corSem  & \corRou  & \blkRou & \fprLen  &  \fprTim &  \fprSem & \fprRou & \fprCon  \\

\hline

%\normalsize{\textbf{\underline{Mimicry}}} \\
%\cline{1-2}

%%%%%%%%%%%%%% Mimicry %%%%%%%%%%%%%%%%%
%\rowcolor{mygray}
\parbox[t]{2mm}{\multirow{18}{*}{\rotatebox[origin=c]{90}{\normalsize{\textbf{Mimicry}}}}} & \mesg-stream-encrypt & \ref{mimicry} &  & &  &  & \lobf & & &  & \si \enc \\
\cline{2-12}

 & \foe & \ref{flow-resist} & \si & & \trn & \trn &  & & \cobf & \trn & \cobf \\
\cline{2-12}

 & \mailmyweb & \ref{flow-resist} & \si &  & \trn & \trn &  &  & \cobf & \trn & \cobf \\
\cline{2-12}

 & Traffic Morphing & \ref{mimicry}  &  &  &  &  & \lobf &  &  &  & \\
\cline{2-12}

& \transteg & \ref{flow-resist} & \si \enc &  &  &  &  &  &  &  & \cobf \\
\cline{2-12}

& \dust & \ref{mimicry} & &  &  & \lobf &  &  &  &  & \si \enc \\
\cline{2-12}

& \skypemorph & \ref{flow-resist} & \si \enc &  & \trn & \trn &  &  &  & \trn & \cobf\\
\cline{2-12}

& \stegotorus & \ref{flow-resist} & \si \enc &  &  &  & \lobf & \tobf & &  & \mux \cobf \\
\cline{2-12}

& \obfsTwo &  \ref{flow-resist} & &  &  &  &  &  &  & & \si \enc \\
\cline{2-12}

& \obfsThree & \ref{flow-resist} &  \si \enc &  &  &  &  &  & \si &  & \si \enc\\
\cline{2-12}

& \obfsFour & \ref{comp-resist}  & \si \enc &  &  &  & \tobf & \lobf & \si \enc & & \si \enc\\
\cline{2-12}

& \sweet & \ref{flow-resist} & \si \enc &  & \trn  & \trn &  &  &  &  \trn & \cobf\\
\cline{2-12}

& \fte & \ref{flow-resist} & \enc &  &  &  &  &  & &  & \enc \cobf\\
\cline{2-12}

& \trist & \ref{flow-resist} & &  &  &  &  &  &  & & \cobf\\
\cline{2-12}

& \rook & \ref{flow-resist}  &  \si \enc  &  &  &  &  & \trn & \trn & \trn & \si \cobf \trn\\
\cline{2-12}

& \castle & \ref{flow-resist}  &  \si \enc  &  & \trn & \trn & \trn & \tobf & \si \trn & \si \trn\\
\cline{2-12}

& \marionette & \ref{flow-resist}  &  \si \enc  &  &  &  & \lobf & \tobf & \cobf & \si \enc\\
\cline{2-12}

& \jumpbox & \ref{flow-resist}  &   &   &  &  & \trn & \trn & \trn & & \\

\noalign{\hrule height 1pt}
%%%%%%%%%%%%% Monitor Driven %%%%%%%%%%%%%%%%
\parbox[t]{2mm}{\multirow{2}{*}{\rotatebox[origin=c]{90}{\normalsize{\textbf{MDFT}}}}} & Khattak & \ref{monitor-driven} & &  &  &  &  &  &  & & \cobf \trn\\
\cline{2-12}

& \gohop & \ref{monitor-driven}  &  \si \enc  &  &  &  & \lobf &  & &  & \si \enc \trn\\

\noalign{\hrule height 1pt}
%%%%%%%%%%%%% Protocol Fingerprinting %%%%%%%%%%%%%%%

\parbox[t]{2mm}{\multirow{4}{*}{\rotatebox[origin=c]{90}{\normalsize{\textbf{Prot. Fingprt.}}}}} & \silentknock & \ref{protocol-fingerprint} &  &  &  &  &  & & \si \cobf \trn & & \\
\cline{2-12}

& \spator & \ref{protocol-fingerprint} & &  &  &  &  &  & \si \cobf \trn &   \\
\cline{2-12}

& \bridgespa & \ref{protocol-fingerprint} & &  &  &  &  &  & \si \cobf \trn &   \\
\cline{2-12}

& \keyspacehopping & \ref{protocol-fingerprint}  &  &  &  &  &  &  & \si \enc &  & \\

\noalign{\hrule height 1pt}
%%%%%%%%%%%%%%% IP Filtering %%%%%%%%%%%%%%%

\parbox[t]{2mm}{\multirow{12}{*}{\rotatebox[origin=c]{90}{\normalsize{\textbf{IP Filtering}}}}} & \cirripede & \ref{ip-resist} & \si \enc &  & \si \enc \trn & \si \enc \trn &  &  & & \si \enc \trn &\si \enc \\
\cline{2-12}

 & \curveball & \ref{ip-resist} & \si \enc &  & \si \enc \trn & \si \enc \trn &  &  & & \si \enc \trn &\si \enc \\
\cline{2-12}

& \telex & \ref{ip-resist} & \si \enc &  & \si \enc \trn & \si \enc \trn &  &  &  &\si \enc \trn  & \si \enc \\
\cline{2-12}

& \tapdance & \ref{ip-resist} & \si \enc & & \si \enc \trn & \si \enc \trn &  &  &  & \si \enc \trn & \si \enc\\
\cline{2-12}

& \defiance & \ref{ip-resist} & &  & \si \enc \cobf  & \si \enc \cobf &  &  &  & \si \enc \cobf & \\
\cline{2-12}

& \flashproxy & \ref{ip-resist} &  &  & \si \trn & \si \trn &  &  &  & \si \trn & \si\\
\cline{2-12}

& \oss & \ref{ip-resist} & \si \enc & & \si \enc \trn & \si \enc \trn &  &  & & \si \enc \trn & \cobf\\
\cline{2-12}

& \miab & \ref{ip-resist} & \si \enc &  & \si & \si &  &  & & \si & \si \enc\\
\cline{2-12}

& \id-based-crypto-tagging & \ref{ip-resist} & \si \enc &  &  &  &  &  & &  &\si \enc\\
\cline{2-12}

& \meek & \ref{ip-resist} &  \si \enc &  & \si \enc \trn & \si \enc \trn &  &  & & \si \enc \trn & \si \enc \\ 
\cline{2-12}

& \cloudtransport & \ref{ip-resist} & \si \enc &  & \si \trn & \si \trn &  &  & & \si \trn & \si \enc \\
\cline{2-12}

& \vpngate & \ref{ip-resist}  &  \si \enc &  & \trn & \trn &  &  & & \trn & \si \enc\\

\noalign{\hrule height 1pt}
%%%%%%%%%%%%%% Composite %%%%%%%%%%%%%%%%

\parbox[t]{2mm}{\multirow{5}{*}{\rotatebox[origin=c]{90}{\normalsize{\textbf{Composite}}}}} & \infranet & \ref{comp-resist} & \si \enc & & \trn & \trn & & & & \trn & \cobf\\
\cline{2-12}

 & \collage & \ref{comp-resist} & \si \enc &  & \trn & \trn &  &  &  & \trn & \si \enc \cobf  \\
\cline{2-12}

& \censorspoofer & \ref{comp-resist} & \si \enc &  & \si \trn & \si \trn &  &  & & \si \trn & \si \enc \\
\cline{2-12}

& \freewave & \ref{comp-resist} & \si \enc &  &\si  & \si &  &  & & \si & \si \enc\\
\cline{2-12}

& \scramblesuit & \ref{comp-resist}  &  \si \enc &  &  &  & \lobf & \tobf & \si \enc &  &  \si \enc \\

\end{tabular} } }
%\hspace*{-1cm}
\end{table*}
%\end{landscape}
%==============End Table============

%% file: conclusion.tex
\section{Conclusion} \label{sec:conclusion}

As censorship attempts have concentrated on disrupting the link
between a censored client and server, a multitude of censorship
resistance schemes have been built to bypass such blocks which we call
link circumvention schemes (\pts). We bring clarity to the complex and
rapidly evolving field of link circumvention by conducting a
systematic survey of \pts in terms of the threat-model defended
against and their evaluation in terms of an abstract evaluation stack.
We have highlighted open challenges in the area of \pts with respect
to common evaluation criteria and modular/flexible development of
systems. These insights have motivated the case for a new framework to
efficiently compose \pts out of reusable components following the
evaluation stack presented in this paper.

%% file: paper.bbl
\begin{thebibliography}{10}

\bibitem{sadia2015hotpets}
S.~Afroz, D.~Fifield, M.~C. Tschantz, V.~Paxson, and J.~D. Tygar, ``{Censorship
  Arms Race: Research vs. Practice},'' in {\em {Workshop on Hot Topics in
  Privacy Enhancing Technologies}}, 2015.

\bibitem{pt_spec_tor}
{Tor}, ``\url{https://www.torproject.org/docs/pluggable-transports.html.en}.''
  Online.
\newblock August 2015.

\bibitem{elahi2012cordon}
T.~Elahi and I.~Goldberg, ``Cordon--a taxonomy of {I}nternet censorship
  resistance strategies,''

\bibitem{Tschantz2014}
M.~C. Tschantz, S.~Afroz, V.~Paxson, and J.~D. Tygar, ``{On Modeling the Costs
  of Censorship},'' arXiv 1409.3211, 2014.

\bibitem{kopsell2004achieve}
S.~K{\"o}psell and U.~Hillig, ``How to achieve blocking resistance for existing
  systems enabling anonymous web surfing,'' in {\em Proceedings of the 2004 ACM
  workshop on Privacy in the electronic society}, pp.~47--58, ACM, 2004.

\bibitem{perng2005censorship}
G.~Perng, M.~K. Reiter, and C.~Wang, ``Censorship resistance revisited,'' in
  {\em Information Hiding}, pp.~62--76, Springer, 2005.

\bibitem{leberknight2012taxonomy}
C.~S. Leberknight, M.~Chiang, H.~V. Poor, and F.~Wong, ``A taxonomy of
  {I}nternet censorship and anti-censorship,'' 2012.

\bibitem{Tor}
{Tor}. Online, November 2015.
\newblock \url{https://www.torproject.org}.

\bibitem{greendam}
{OpenNet Initiative}, ``{China's Green Dam: The Implications of Government
  Control Encroaching on the Home PC}.''
  {https://opennet.net/chinas-green-dam-the-implications-government-control-encroaching-home-pc}.

\bibitem{TomSkype}
J.~Knockel, J.~R. Crandall, and J.~Saia, ``{Three Researchers, Five
  Conjectures: An Empirical Analysis of TOM-Skype Censorship and
  Surveillance},'' in {\em Free and Open Communications on the {I}nternet},
  (San Francisco, CA, USA), USENIX, 2011.

\bibitem{Bamman12}
{D. Bamman and B. O'Connor and N. Smith}, ``{Censorship and deletion practices
  in Chinese social media}.'' Online, November 2015.
\newblock
  \url{http://journals.uic.edu/ojs/index.php/fm/article/view/3943/3169}.

\bibitem{Zhu2013}
T.~Zhu, D.~Phipps, A.~Pridgen, J.~R. Crandall, and D.~S. Wallach, ``The
  velocity of censorship: High-fidelity detection of microblog post
  deletions,'' in {\em Proceedings of the 22nd USENIX Conference on Security},
  SEC'13, (Berkeley, CA, USA), pp.~227--240, USENIX Association, 2013.

\bibitem{Weiboscope}
{Journalism and Media Studies Centre}, ``Weiboscope.'' Online, November 2015.
\newblock \url{http://weiboscope.jmsc.hku.hk}.

\bibitem{Anderson2013}
C.~Anderson, ``{Dimming the {I}nternet: Detecting Throttling as a Mechanism of
  Censorship in Iran},'' tech. rep., University of Pennsylvania, 2013.

\bibitem{gfw-review:httpurl}
``{{HTTP} {URL}/keyword detection in depth}.''
  {http://gfwrev.blogspot.jp/2010/03/http-url.html}, 2010.

\bibitem{Winter2012}
P.~Winter and S.~Lindskog, ``How the great firewall of {C}hina is blocking
  {T}or,'' in {\em Free and Open Communications on the {I}nternet}, (Bellevue,
  WA, USA), USENIX, 2012.

\bibitem{egypt-outage}
{Renesys},
  ``\url{http://research.dyn.com/2011/01/egypt-leaves-the-internet/}.'' Online.
\newblock August 2015.

\bibitem{libya-outage}
{Renesys},
  ``\url{http://research.dyn.com/2011/08/the-battle-for-tripolis-intern/}.''
  Online.
\newblock August 2015.

\bibitem{sudan-outage}
{Renesys}, ``\url{http://research.dyn.com/2013/09/internet-blackout-sudan/}.''
  Online.
\newblock August 2015.

\bibitem{myanmar-outage}
{Renesys}, ``\url{http://research.dyn.com/2013/08/myanmar-internet/}.'' Online.
\newblock August 2015.

\bibitem{Anonymous12}
Anonymous, ``{The Collateral Damage of {I}nternet Censorship by DNS
  Injection},'' {\em SIGCOMM Comput. Commun. Rev.}, vol.~42, pp.~21--27, June
  2012.

\bibitem{Clayton2006}
R.~Clayton, ``{Failures in a Hybrid Content Blocking System},'' in {\em Privacy
  Enhancing Technologies}, (Cambridge, England), pp.~78--92, Springer, 2006.

\bibitem{l7-filter}
{l7-filter}, ``\url{http://research.dyn.com/2013/08/myanmar-internet/}.''
  Online.
\newblock August 2015.

\bibitem{bro}
{Bro}, ``\url{https://www.bro.org}.'' Online.
\newblock August 2015.

\bibitem{snort}
{Snort}, ``\url{https://www.snort.org}.'' Online.
\newblock August 2015.

\bibitem{nDPI}
{nDPI}, ``\url{http://www.ntop.org/products/ndpi/}.'' Online.
\newblock August 2015.

\bibitem{Dainotti2012}
A.~Dainotti, A.~Pescap\'{e}, and K.~Claffy, ``{Issues and future directions in
  traffic classification},'' {\em IEEE Network}, vol.~26, pp.~35--40, Jan 2012.

\bibitem{Sommer2010}
R.~Sommer and V.~Paxson, ``Outside the closed world: On using machine learning
  for network intrusion detection,'' in {\em In Proceedings of the IEEE
  Symposium on Security and Privacy}, 2010.

\bibitem{Dorfinger2011}
P.~Dorfinger, G.~Panholzer, and W.~John, ``Entropy estimation for real-time
  encrypted traffic identification (short paper),'' in {\em Traffic Monitoring
  and Analysis} (J.~Domingo-Pascual, Y.~Shavitt, and S.~Uhlig, eds.), vol.~6613
  of {\em Lecture Notes in Computer Science}, pp.~164--171, Springer Berlin
  Heidelberg, 2011.

\bibitem{Bernaille2007}
L.~Bernaille and R.~Teixeira, ``Early recognition of encrypted applications,''
  in {\em Proceedings of the 8th International Conference on Passive and Active
  Network Measurement}, PAM'07, (Berlin, Heidelberg), pp.~165--175,
  Springer-Verlag, 2007.

\bibitem{Wright2006}
C.~V. Wright, F.~Monrose, and G.~M. Masson, ``On inferring application protocol
  behaviors in encrypted network traffic,'' {\em J. Mach. Learn. Res.}, vol.~7,
  pp.~2745--2769, Dec. 2006.

\bibitem{Wiley2011}
B.~Wiley, ``Blocking-resistant protocol classification using {B}ayesian model
  selection,'' tech. rep., University of Texas at Austin, 2011.

\bibitem{dust}
B.~Wiley, ``Dust: A blocking-resistant {I}nternet transport protocol,'' tech.
  rep., School of Information, University of Texas at Austin, 2011.

\bibitem{obfs-openssh}
B.~Leidl, ``{obfuscated-openssh}.''
  {https://github.com/brl/obfuscated-openssh/blob/master/README.obfuscation},
  2010.

\bibitem{panchenko2011}
A.~Panchenko, L.~Niessen, A.~Zinnen, and T.~Engel, ``Website fingerprinting in
  onion routing based anonymization networks,'' in {\em Proceedings of the
  Workshop on Privacy in the Electronic Society (WPES 2011)}, ACM, October
  2011.

\bibitem{Sun2002}
Q.~Sun, D.~R. Simon, Y.-M. Wang, W.~Russell, V.~N. Padmanabhan, and L.~Qiu,
  ``Statistical identification of encrypted web browsing traffic,'' in {\em
  Proceedings of the 2002 IEEE Symposium on Security and Privacy}, May 2002.

\bibitem{Hintz2002}
A.~Hintz, ``Fingerprinting websites using traffic analysis,'' in {\em
  Proceedings of Privacy Enhancing Technologies workshop (PET 2002)}
  (R.~Dingledine and P.~Syverson, eds.), Springer-Verlag, LNCS 2482, April
  2002.

\bibitem{Bissias2005}
G.~D. Bissias, M.~Liberatore, and B.~N. Levine, ``Privacy vulnerabilities in
  encrypted {HTTP} streams,'' in {\em Proceedings of Privacy Enhancing
  Technologies workshop (PET 2005)}, pp.~1--11, May 2005.

\bibitem{Wilde2012}
{T. Wilde}, ``Great firewall {Tor} probing.'' Online, November 2015.
\newblock \url{https://gist.github.com/da3c7a9af01d74cd7de7}.

\bibitem{lantern}
Lantern, ``\url{https://getlantern.org}.'' Online.
\newblock August 2015.

\bibitem{Psiphon}
{Psiphon Inc.}, ``{Psiphon}.'' Online, November 2015.
\newblock \url{https://psiphon.ca/en/index.html}.

\bibitem{vpngate:2014}
D.~Nobori and Y.~Shinjo, ``{VPN} {G}ate: A volunteer-organized public {VPN}
  relay system with blocking resistance for bypassing government censorship
  firewalls,'' in {\em Networked Systems Design and Implementation}, USENIX,
  2014.

\bibitem{meek}
meek, ``\url{https://trac.torproject.org/projects/tor/wiki/doc/meek}.'' Online.
\newblock August 2015.

\bibitem{domain_fronting:2015}
D.~Fifield, C.~Lan, R.~Hynes, P.~Wegmann, and V.~Paxson, ``Blocking-resistant
  communication through domain fronting,'' {\em Proceedings on Privacy
  Enhancing Technologies}, vol.~2015, no.~2, 2015.

\bibitem{oss}
D.~Fifield, G.~Nakibly, and D.~Boneh, ``{OSS}: Using online scanning services
  for censorship circumvention,'' in {\em Privacy Enhancing Technologies},
  vol.~7981 of {\em {LNCS}}, pp.~185--204, 2013.

\bibitem{pdfmyurl}
pdfmyurl, ``\url{http://pdfmyurl.com}.'' Online.
\newblock August 2015.

\bibitem{cloud-transport}
C.~MiscBrubaker, A.~Houmansadr, and V.~Shmatikov, ``{CloudTransport: Using
  Cloud Storage for Censorship-Resistant Networking},'' in {\em Privacy
  Enhancing Technologies Symposium}, Springer, 2014.

\bibitem{collage}
S.~Burnett, N.~Feamster, and S.~Vempala, ``{Chipping Away at Censorship
  Firewalls with User-Generated Content},'' in {\em USENIX Security Symposium},
  (Washington, DC, USA), USENIX, 2010.

\bibitem{miab}
L.~Invernizzi, C.~Kruegel, and G.~Vigna, ``{Message In A Bottle: Sailing Past
  Censorship},'' in {\em Annual Computer Security Applications Conference},
  (New Orleans, LA, USA), ACM, 2013.

\bibitem{defiance}
P.~Lincoln, I.~Mason, P.~Porras, V.~Yegneswaran, Z.~Weinberg, J.~Massar, W.~A.
  Simpson, P.~Vixie, and D.~Boneh, ``Bootstrapping communications into an
  anti-censorship system,'' in {\em Proceedings of the USENIX Workshop on Free
  and Open Communications on the {I}nternet (FOCI 2012)}, August 2012.

\bibitem{optimizing_proxy_placement}
J.~Cesareo, J.~Karlin, M.~Schapira, and J.~Rexford, ``Optimizing the placement
  of implicit proxies,'' tech. rep., Department of Computer Science, Princeton
  University, Jun 2012.

\bibitem{routing_around_decoys}
M.~Schuchard, J.~Geddes, C.~Thompson, and N.~Hopper, ``Routing around decoys,''
  in {\em Proceedings of the 2012 ACM Conference on Computer and Communications
  Security}, CCS '12, (New York, NY, USA), pp.~85--96, ACM, 2012.

\bibitem{decoy_routing_costs}
A.~Houmansadr, E.~L. Wong, and V.~Shmatikov, ``{No Direction Home: The True
  Cost of Routing Around Decoys},'' in {\em Network and Distributed System
  Security Symposium (NDSS)}, 2014.

\bibitem{telex}
E.~Wustrow, S.~Wolchok, I.~Goldberg, and J.~A. Halderman, ``{Telex:
  Anticensorship in the Network Infrastructure},'' in {\em USENIX Security
  Symposium}, (San Francisco, CA, USA), USENIX, 2011.

\bibitem{tapdance}
E.~Wustrow, C.~M. Swanson, and J.~A. Halderman, ``Tapdance: End-to-middle
  anticensorship without flow blocking,'' in {\em 23rd USENIX Security
  Symposium (USENIX Security 14)}, (San Diego, CA), pp.~159--174, USENIX
  Association, Aug. 2014.

\bibitem{foci11-decoy}
J.~Karlin, D.~Ellard, A.~W. Jackson, C.~E. Jones, G.~Lauer, D.~P. Mankins, and
  W.~T. Strayer, ``Decoy routing: Toward unblockable {I}nternet
  communication,'' in {\em Proceedings of the USENIX Workshop on Free and Open
  Communications on the {I}nternet (FOCI 2011)}, August 2011.

\bibitem{curveball}
J.~Karlin, D.~Ellard, A.~W. Jackson, C.~E. Jones, G.~Lauer, D.~P. Mankins, and
  W.~T. Strayer, ``{Decoy Routing: Toward Unblockable {I}nternet
  Communication},'' in {\em Free and Open Communications on the {I}nternet},
  (San Francisco, CA, USA), USENIX, 2011.

\bibitem{id-based-crypto-tagging}
T.~Ruffing, J.~Schneider, and A.~Kate, ``Identity-based steganography and its
  applications to censorship resistance,'' 2013.
\newblock 6th Workshop on Hot Topics in Privacy Enhancing Technologies (HotPETs
  2013).

\bibitem{Khattak2013}
S.~Khattak, M.~Javed, P.~D. Anderson, and V.~Paxson, ``{Towards Illuminating a
  Censorship Monitor's Model to Facilitate Evasion},'' in {\em Free and Open
  Communications on the {I}nternet}, (Washington, DC, USA), USENIX, 2013.

\bibitem{silentknock}
E.~Y. Vasserman, N.~Hopper, and J.~Tyra, ``Silent knock : practical, provably
  undetectable authentication.,'' {\em Int. J. Inf. Sec.}, vol.~8, no.~2,
  pp.~121--135, 2009.

\bibitem{bridgespa}
R.~Smits, D.~Jain, S.~Pidcock, I.~Goldberg, and U.~Hengartner, ``{BridgeSPA}:
  Improving {Tor} bridges with single packet authorization,'' in {\em
  Proceedings of the 10th Annual ACM Workshop on Privacy in the Electronic
  Society}, WPES '11, (New York, NY, USA), pp.~93--102, ACM, 2011.

\bibitem{spator}
R.~Smits, D.~Jain, S.~Pidcock, I.~Goldberg, and U.~Hengartner, ``{SPATor}:
  Improving {Tor} bridges with single packet authorization,''

\bibitem{keyspace_hopping:2003}
N.~Feamster, M.~Balazinska, W.~Wang, H.~Balakrishnan, and D.~Karger,
  ``Thwarting web censorship with untrusted messenger discovery,'' in {\em
  Privacy Enhancing Technologies}, pp.~125--140, Springer, 2003.

\bibitem{foe}
{foe-project}, ``\url{https://code.google.com/p/foe-project/}.'' Online.
\newblock August 2015.

\bibitem{mailmyweb}
{MailMyWeb}, ``\url{http://www.mailmyweb.com}.'' Online.
\newblock August 2015.

\bibitem{sweet}
W.~Zhou, A.~Houmansadr, M.~Caesar, and N.~Borisov, ``{SWEET: Serving the Web by
  Exploiting Email Tunnels},'' in {\em Hot Topics in Privacy Enhancing
  Technologies}, (Bloomington, IN, USA), Springer, 2013.

\bibitem{skypemorph}
H.~Mohajeri~Moghaddam, B.~Li, M.~Derakhshani, and I.~Goldberg, ``Skypemorph:
  Protocol obfuscation for {Tor} bridges,'' in {\em Proceedings of the 2012 ACM
  Conference on Computer and Communications Security}, CCS '12, (New York, NY,
  USA), pp.~97--108, ACM, 2012.

\bibitem{transteg}
W.~Mazurczyk, P.~Szaga, and K.~Szczypiorski, ``{Using Transcoding for Hidden
  Communication in IP Telephony},'' 2011.

\bibitem{fte}
K.~P. Dyer, S.~E. Coull, T.~Ristenpart, and T.~Shrimpton, ``Protocol
  misidentification made easy with format-transforming encryption,'' in {\em
  Proceedings of the 20th ACM conference on Computer and Communications
  Security (CCS 2013)}, November 2013.

\bibitem{parrot}
A.~Houmansadr, C.~Brubaker, and V.~Shmatikov, ``{The Parrot is Dead: Observing
  Unobservable Network Communications},'' in {\em Symposium on Security \&
  Privacy}, (San Francisco, CA, USA), IEEE, 2013.

\bibitem{Geddes2013}
J.~Geddes, M.~Schuchard, and N.~Hopper, ``{Cover Your ACKs: Pitfalls of Covert
  Channel Censorship Circumvention},'' in {\em Computer and Communications
  Security}, (Berlin, Germany), ACM, 2013.

\bibitem{facet}
S.~Li, M.~Schliep, and N.~Hopper, ``Facet: Streaming over videoconferencing for
  censorship circumvention,'' in {\em Proceedings of the 13th Workshop on
  Privacy in the Electronic Society}, WPES '14, (New York, NY, USA),
  pp.~163--172, ACM, 2014.

\bibitem{freewave}
A.~Houmansadr, T.~Riedl, N.~Borisov, and A.~Singer, ``{I Want my Voice to be
  Heard: IP over Voice-over-IP for Unobservable Censorship Circumvention},'' in
  {\em Proceedings of the Network and Distributed System Security Symposium -
  {NDSS}'13}, {I}nternet Society, February 2013.

\bibitem{jumpbox:2014}
J.~Massar, I.~Mason, L.~Briesemeister, and V.~Yegneswaran, ``Jumpbox--a
  seamless browser proxy for {Tor} pluggable transports,'' {\em Security and
  Privacy in Communication Networks. Springer}, p.~116, 2014.

\bibitem{castle_game:2015}
B.~Hahn, R.~Nithyanand, P.~Gill, and R.~Johnson, ``Games without frontiers:
  Investigating video games as a covert channel,'' {\em CoRR},
  vol.~abs/1503.05904, 2015.

\bibitem{rook_video:2015}
P.~Vines and T.~Kohno, ``Rook: Using video games as a low-bandwidth censorship
  resistant communication platform,'' Tech. Rep. UW-CSE-2015-03-03, University
  of Washington, Mar 2015.

\bibitem{marionette:2015}
K.~P. Dyer, S.~E. Coull, and T.~Shrimpton, ``Marionette: A programmable network
  traffic obfuscation system,'' in {\em 24th USENIX Security Symposium (USENIX
  Security 15)}, (Washington, D.C.), pp.~367--382, USENIX Association, Aug.
  2015.

\bibitem{scramblesuit}
P.~Winter, T.~Pulls, and J.~Fuss, ``{ScrambleSuit: A Polymorphic Network
  Protocol to Circumvent Censorship},'' in {\em Workshop on Privacy in the
  Electronic Society}, (Berlin, Germany), ACM, 2013.

\bibitem{mesg-stream-encrypt}
J.~Salowey, H.~Zhou, P.~Eronen, and H.~Tschofenig., ``\url{Message Stream
  Encryption}.'' {http://tinyurl.com/m9ml7ct}, 2006.

\bibitem{obfs2}
{Tor},
  ``\url{https://gitweb.torproject.org/pluggable-transports/obfsproxy.git/blob/HEAD:/doc/obfs2/obfs2-protocol-spec.txt}.''
  Online.
\newblock August 2015.

\bibitem{obfs3}
{Tor},
  ``\url{https://gitweb.torproject.org/pluggable-transports/obfsproxy.git/blob/HEAD:/doc/obfs3/obfs3-protocol-spec.txt}.''
  Online.
\newblock August 2015.

\bibitem{obfs4}
{Tor},
  ``\url{https://github.com/Yawning/obfs4/blob/5bdc376e2abaf5ac87816b763f5b26e314ee9536/doc/obfs4-spec.txt}.''
  Online.
\newblock August 2015.

\bibitem{ntor_academic_paper:2013}
I.~Goldberg, D.~Stebila, and B.~Ustaoglu, ``Anonymity and one-way
  authentication in key exchange protocols,'' {\em Designs, Codes and
  Cryptography}, vol.~67, no.~2, pp.~245--269, 2013.

\bibitem{ntor_specs}
{Nick Mathewson},
  ``\url{https://gitweb.torproject.org/torspec.git/blob/HEAD:/proposals/216-ntor-handshake.txt}.''
  Online.
\newblock August 2015.

\bibitem{gohop:2014}
Y.~Wang, P.~Ji, B.~Ye, P.~Wang, R.~Luo, and H.~Yang, ``Go{H}op: Personal {VPN}
  to defend from censorship,'' in {\em International Conference on Advanced
  Communication Technology}, IEEE, 2014.

\bibitem{clayton:pet2006}
R.~Clayton, S.~J. Murdoch, and R.~N.~M. Watson, ``{Ignoring the Great Firewall
  of China},'' in {\em Proceedings of the Sixth Workshop on Privacy Enhancing
  Technologies (PET 2006)} (G.~Danezis and P.~Golle, eds.), pp.~20--35,
  Springer, June 2006.

\bibitem{westChamberS1}
``{Scholar Zhang: Intrusion detection evasion and black box mechanism research
  of the Great Firewall of China}.'' {https://code.google.com/p/scholarzhang/},
  2010.

\bibitem{westChamberS2}
``west-chamber-season-2.'' {https://code.google.com/p/west-chamber-season-2/},
  2010.

\bibitem{westChamberS3}
``west-chamber-season-3.'' {https://github.com/liruqi/west-chamber-season-3/},
  2011.

\bibitem{infranet}
N.~Feamster, M.~Balazinska, G.~Harfst, H.~Balakrishnan, and D.~R. Karger,
  ``Infranet: Circumventing web censorship and surveillance.,'' in {\em USENIX
  Security Symposium} (D.~Boneh, ed.), pp.~247--262, USENIX, 2002.

\bibitem{AdversaryLab}
{Brandon Wiley}, ``{AdversaryLab}.'' Online, November 2015.
\newblock \url{https://github.com/blanu/AdversaryLab/}.

\bibitem{david2015hotpets}
D.~Fifield, L.~N. Lee, S.~Egelman, and D.~Wagner, ``{Tor's Usability for
  Censorship Circumvention},'' in {\em {Workshop on Hot Topics in Privacy
  Enhancing Technologies}}, 2015.

\bibitem{ElahiMG16}
T.~Elahi, J.~A. Doucette, H.~Hosseini, S.~J. Murdoch, and I.~Goldberg, ``A
  framework for the game-theoretic analysis of censorship resistance,'' {\em
  Proceedings on Privacy Enhancing Technologies}, vol.~2016, July 2016.

\bibitem{Fog}
{The Tor Project}, ``{Fog}.'' Online, November 2015.
\newblock \url{https://gitweb.torproject.org/pluggable-transports/fog.git}.

\bibitem{tweakable}
{Steven J. Murdoch}, ``{Pluggable Transport Component Architecture}.''
\newblock
  \url{https://gitweb.torproject.org/sjm217/torspec.git/tree/pt-components.txt?h=pt-components}.

\end{thebibliography}
